\documentclass[pre,aps,12pt,superscriptaddress]{revtex4-1}

\usepackage{graphicx}
\usepackage{appendix}
\usepackage{amsmath,amssymb}

\newcommand{\degree}{\ensuremath{^\circ}}
\newcommand{\ignorar}[1]{}

\begin{document}

\title{How reliable are Finite-Size Lyapunov Exponents for the
assessment of ocean dynamics?}

\author{Ismael Hern\'andez-Carrasco} \author{Crist\'obal L\'opez}
\author{Emilio Hern\'andez-Garc\'ia} 

\affiliation{
IFISC, Instituto de F\'isica Interdisciplinar y Sistemas Complejos (CSIC-UIB),
07122 Palma de Mallorca, Spain}
\author{Antonio Turiel}
\affiliation{Institut de Ci\`encies del Mar, CSIC,
Passeig Mar\'{\i}tim de la Barceloneta 37-49, 08003
  Barcelona, Spain}

\date{\today}

\begin{abstract}
Much of atmospheric and oceanic transport is associated with coherent structures.
Lagrangian methods are emerging as optimal tools
for their  identification and analysis.
An important Lagrangian technique which is starting to be widely used in
oceanography is that of Finite-Size Lyapunov Exponents (FSLEs).
Despite this growing relevance there are still many open questions concerning
the reliability of the FSLEs in order to analyse the ocean dynamics.
In particular, it is still unclear how robust they are when confronted with real data.
In this paper we analyze the effect on this Lagrangian technique
of the two most important effects when facing real data,
namely noise and dynamics of unsolved scales.
Our results, using as a benchmarch data from a primitive numerical
model of the Mediterranean Sea, show that even when some dynamics
is missed the FSLEs results still give an accurate picture of the oceanic transport properties.

\end{abstract}

\maketitle

Lagrangian viewpoint
\cite{Griffa.1996, Buffoni.1997, Iudicone.2002, Mancho.2006, Molcard.2006}. Lagrangian
diagnostics exploit the spatio-temporal variability of the
velocity field by following fluid particle trajectories, in
contrast with Eulerian diagnostics, which analyze only frozen
snapshots of data. Among Lagrangian techniques the most used ones
involve the computation of local Lyapunov exponents (LLE) which
measure the relative dispersion of transported particles
\cite{Artale.1997,Aurell.1997,
Haller.1998,Haller.2000,Boffetta.2001,Iudicone.2002}. LLEs give information on
dispersion processes but also, and even more importantly, can be
used to detect and visualize Lagrangian Coherent Structures (LCSs)
in the flow like vortices, barriers to transport, fronts,
etc.\cite{Haller.2000b,Joseph.2002,Koh.2002,Lapeyre.2002,Beron-Vera.2008,
Chaos.2010}.

The standard definition of Lyapunov exponents \cite{Ott.1993}
involves a double limit, in which infinitely-long times and infinitesimal
initial separations are taken. These limits can not be practically
implemented when dealing with realistic flows of geophysical
origin. Over real data, LLEs are defined by relaxing some of the limit
procedures. In finite-time Lyapunov exponents (FTLE)
\cite{Ott.1993,Lapeyre.2002} trajectory separations are computed
starting still at infinitesimal distance, but only for a finite
time. In the case of finite-size Lyapunov exponents (FSLE), the key
tool used in our work
\cite{Aurell.1997,Boffetta.2001}, one computes the time which is taken
for two trajectories, initially separated by a finite distance, to
reach a larger final finite distance.

FSLEs are attracting the attention of the oceanographic community
\cite{Artale.1997,Iudicone.2002,
dOvidio.2004,Molcard.2006,Garcia-Olivares.2007,Haza.2008,dOvidio.2009,
TewKai.2009,Poje.2010}.
The main reasons for this interest, within the framework of their
ability for studying LCSs and dispersion processes, are the
following: a) they identify and display the dynamical structures
in the flow that strongly organizes fluid motion (the above
mentioned LCSs, which are defined as {\it ridges of the 
FSLEs fields}); b) they are relatively easy to compute; c) they
provide extra information on characteristics time-scales for the
dynamics; and d) they are able to reveal oceanic structures below
the nominal resolution of the velocity field being analyzed. In
addition, FSLEs are specially suited to analyze transport in
closed areas \cite{Artale.1997}.

Despite the growing number of applications of FSLEs, a rigorous
analysis of many of their properties is still lacking. There are
two main concerns before applying FSLE to real data, namely the
effect of noise and the role of observation scale. Concerning
noise, real data are discretized and noisy, and this can affect
the reliability of FSLE-based diagnostics
(recent related studies for FTLE can be consulted in \cite{Chaos.2010}).
Concerning scale
properties, FSLEs can be obtained over a grid finer than that of
the data. This enables to study submesoscale processes under the
typical mesoscales (below $10$ kilometers) that nowadays provide,
for example, altimetry data \cite{dOvidio.2009,TewKai.2009}. But
then the question is if finer-grid LCSs are meaningful or just an
artifact. On the other side, we can have access to a
limited-resolution velocity field, and then ask ourselves if any
refinement in the velocity grid (by improved data acquisition, for
instance) is going to modify our previous assessment of LCS at the
rougher scale. The main objective of this paper is to address
these questions, in particular with reference to their potential
applications into ocean dynamics. 
A related study of the sensitivity of relative dispersion
of particles statistics, when the spatial resolution of the velocity 
field changes, has appeared recently \cite{Poje.2010}.

The benchmark for the study of FSLEs properties used in this work
is a the two-dimensional velocity field of the marine
surface obtained from a numerical model of the Mediterranean Sea.
The first half of the paper is devoted to
study what we have just called scale properties of the FSLE field.
By changing gradually the resolution of the grid on which they are
computed,  we will show that they have typical multifractal
properties. This means, in particular, that FSLEs obtained for a
finer resolution than that provided by the data provides
non-artificial information. Subsequently, we will consider a
somewhat opposite case, i.e., what happens to the FSLEs if the
velocity-data grid is changed. Their robustness under
data-resolution transformations will be discussed. The second half
of the work will analyze the effect of noise. Again, two different
scenarios are considered: a) uncertainties in the velocity data,
and b) noise in the particle trajectories. FSLEs will be shown to
be robust against these two sources of error, and the reasons for
this will be discussed.

\section{Description of the data}
\label{Sec:Data}

\begin{figure}[htb]
\begin{center}
\includegraphics[width=10cm]{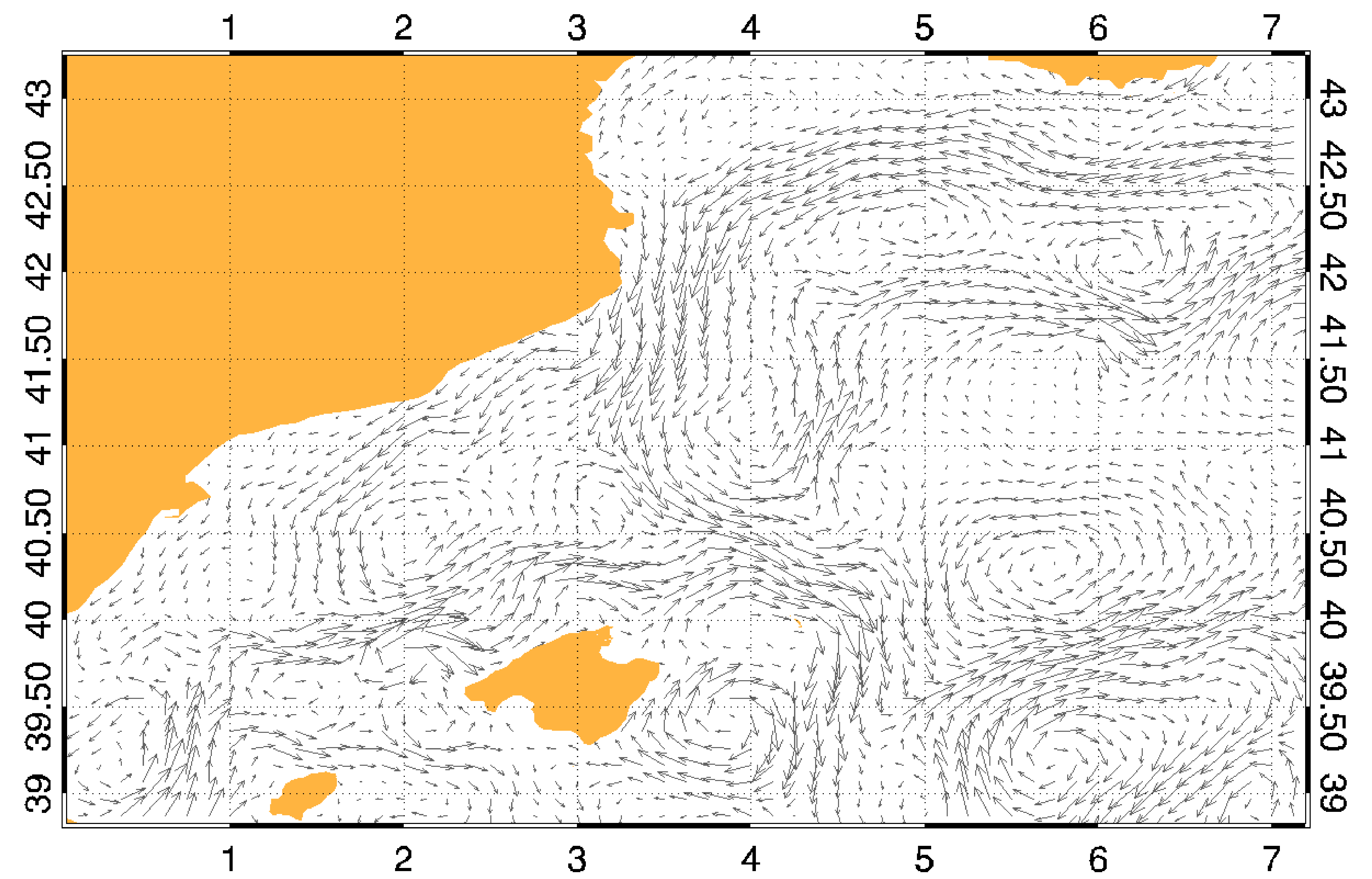}
\end{center}
\caption{Snapshot of the surface velocity field
of the Balearic Sea corresponding to day $640$ in the DieCAST simulation.}
\label{fig:velocity}
\end{figure}

We analyze a velocity dataset generated with the DieCast (Dietrich
for Center Air Sea Technology) numerical ocean model adapted to
the Mediterranean Sea \cite{Fernandez.2005}. The dataset has been
already used in previous Lagrangian studies
\cite{dOvidio.2004,Schneider.2005a,Mancho.2008}. DieCast is a
primitive-equation, z-level, finite difference ocean model which
uses the hydrostatic, incompressible and rigid lid approximations.
At each grid point, horizontal resolution is the same in both the
longitudinal, $\phi$, and latitudinal, $\lambda$, directions, with
resolutions $\Delta \phi = 1/8^o$ and $\Delta \lambda = \Delta
\phi \cos \lambda$. Vertical resolution is variable with 30
layers. Annual climatologic forcing is used, so that it is enough
to keep a temporal resolution of one day. We will use velocity
data corresponding to the second layer, which has its center at a
depth of 16 meters. This sub-surface layer is representative of
the marine surface circulation and is not directly driven by wind.
We have recorded daily velocities for five years, and concentrate
our work in the area of the Balearic Sea. In
Figure~\ref{fig:velocity} we show a snapshot of the velocity field
from the DieCAST model.

\section{Definition and implementation of FSLEs}

FSLEs provide a measure of dispersion as a
function of the spatial resolution, serving to isolate the different
regimes corresponding to different length scales of oceanic flows, as
well as identifying the LCSs present in the data.
To calculate the FSLEs we have to know the trajectories of fluid particles,
which are computed by integrating the equations of motion for which
we need the velocity data $(u,v)$.
 FSLE are computed
from $\tau$, the time required for two particles of fluid (one of them placed
at $(x,y)$) to separate from an initial (at time $t$) distance of
$\delta_0$ to a final distance of $\delta_f$ , as:

\begin{equation}
\Lambda (x,y,t,\delta_0,\delta_f)\; =\; \frac{1}{\tau} \log
\frac{\delta_f}{\delta_0}.
\label{eq:FSLE}
\end{equation}

In principle obtaining $\Lambda (x,y,t,\delta_0,\delta_f)$ would
imply to consider all the trajectories starting from points at
distance $\delta_0$ from our basis point; in practice, when
confronted with regular, discretized grids, only the four closest
neighbors are considered. It is convenient to choose $\delta_0$
to be the intergrid spacing among the points on which the FSLEs
will be computed, i.e., it is the resolution of the ``FSLE grid".
The details of the calculation of the FSLEs are in Appendix \ref{Ap:FSLE}.

\section{Effect of sampling scale on FSLEs}

Notice that we distinguish two types of grids, namely the FSLE
grid (in the following ``F-grid'') and that where the velocity
field is given (called velocity grid or ``V-grid''). These two
grids need not to coincide. We explore first the effect of
changing the resolution of FSLE grid.

\begin{figure}[htb]
 \begin{center}
 \leavevmode
 \includegraphics[width=7.5cm]{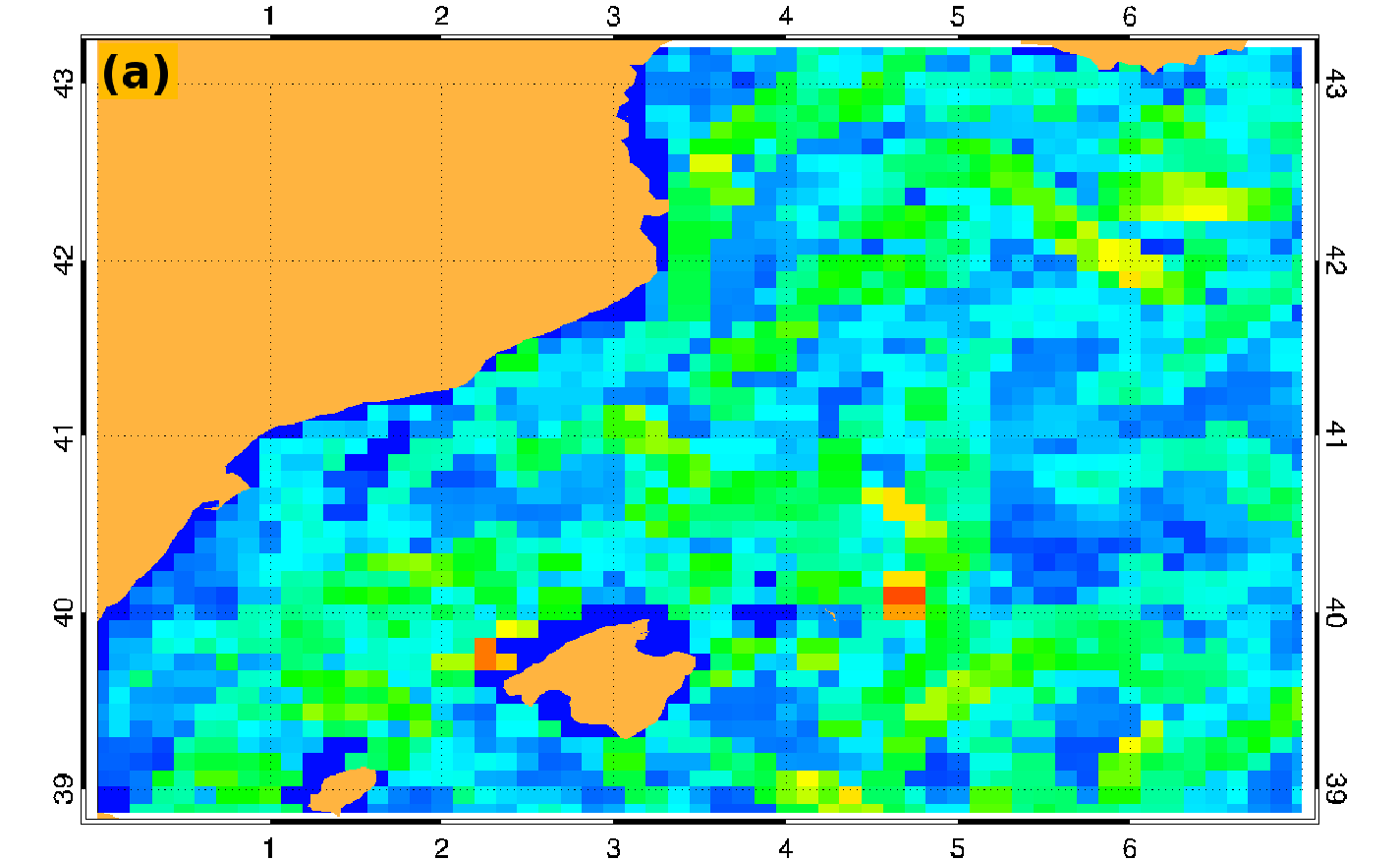}
 \includegraphics[width=7.5cm]{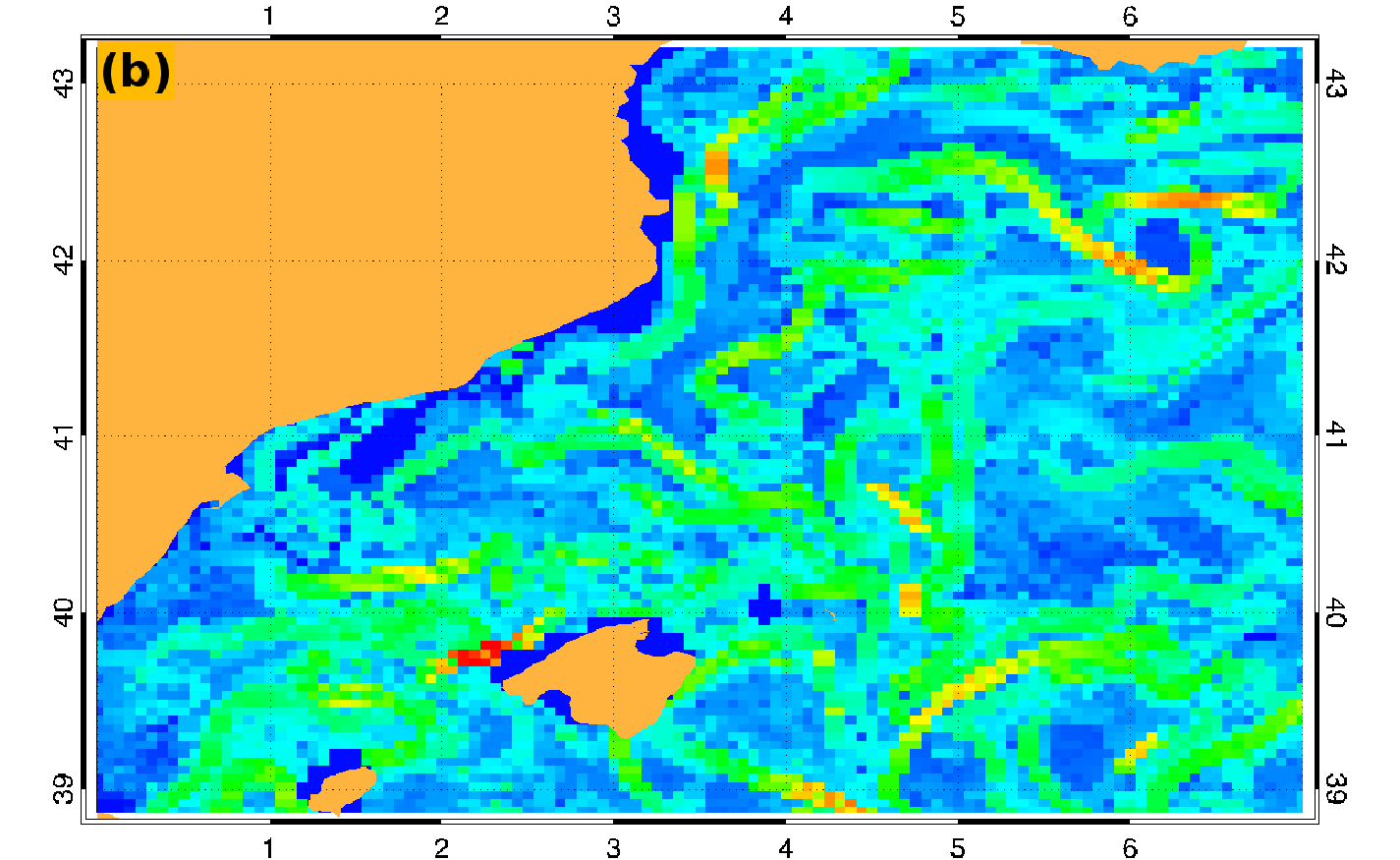}
 \\
 \leavevmode
 \includegraphics[width=7.5cm]{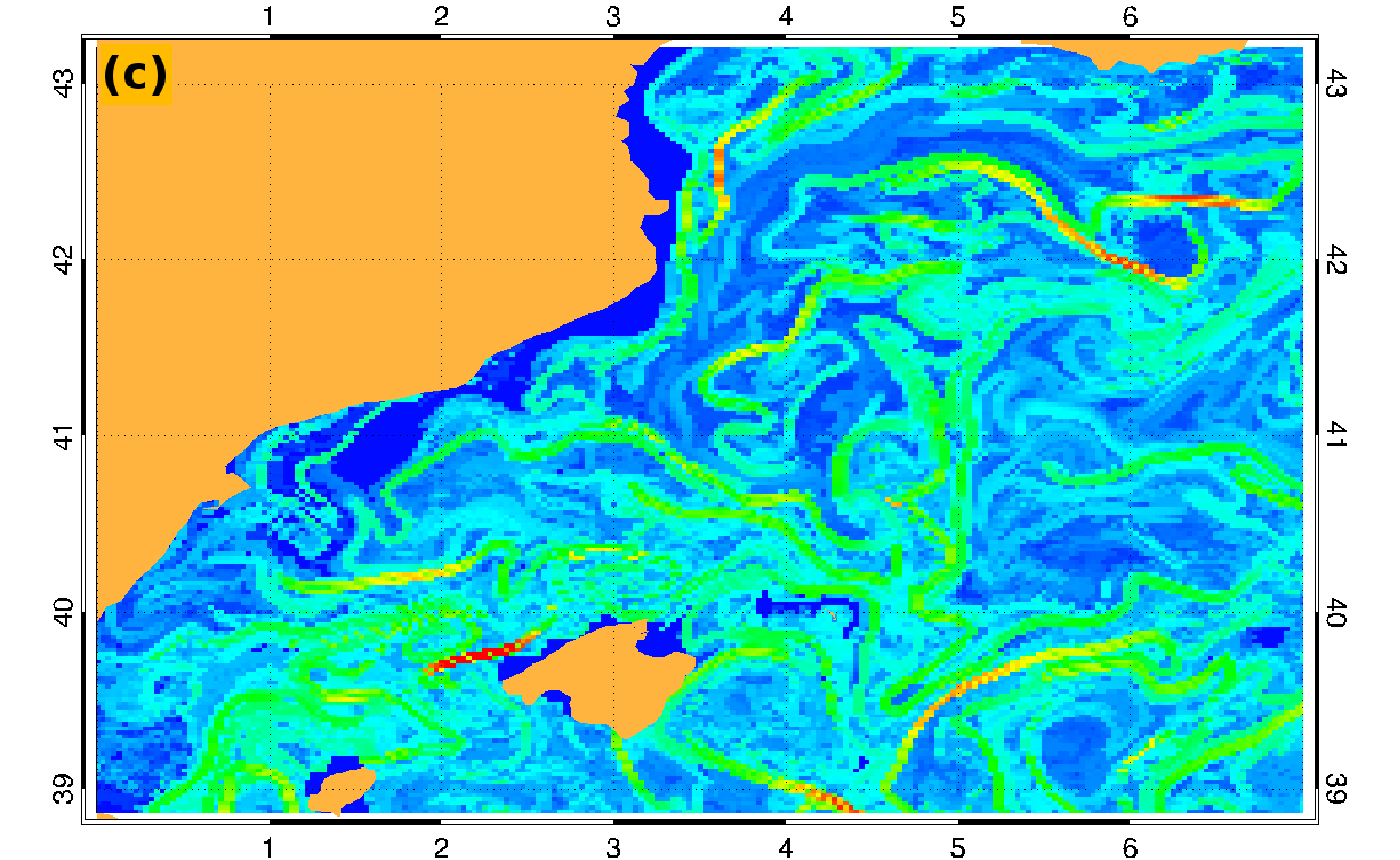}
 \includegraphics[width=7.5cm]{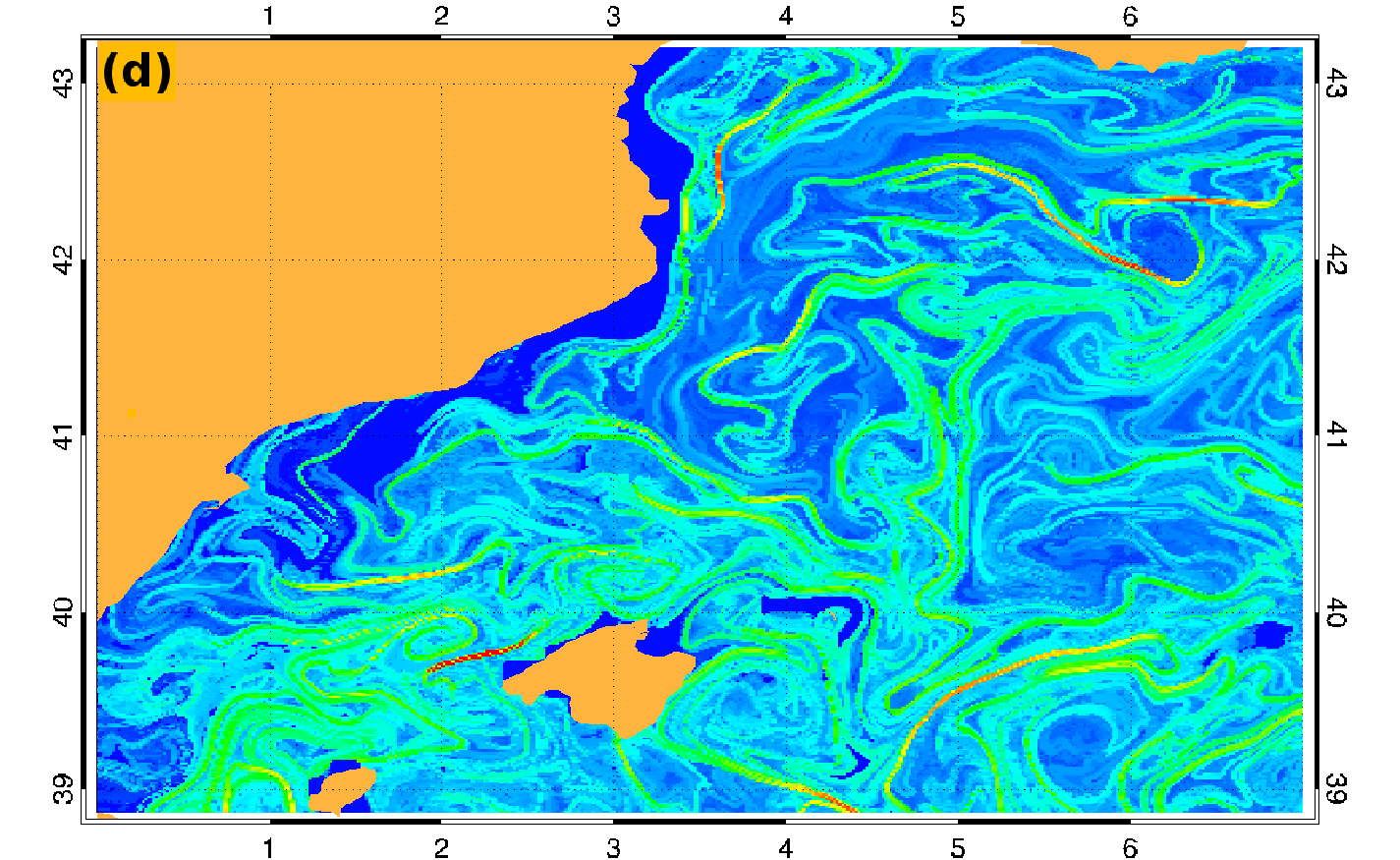}
 \\
 \includegraphics[width=10cm,height=1cm]{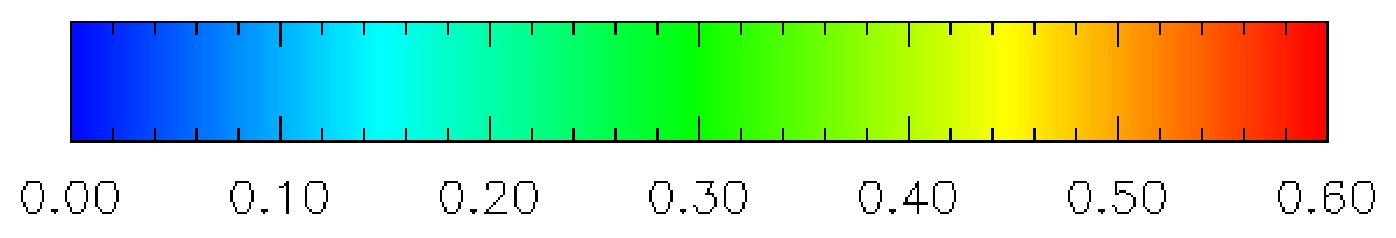}
 \end{center}
 \caption{Snapshots of FSLEs backward in
 time starting from day $640$ at
  different F-grid resolutions: a) $\delta_0=1/8 \degree$, b) $\delta_0=1/16 \degree$,
 c) $\delta_0=1/32 \degree$, d) $\delta_0=1/64 \degree$. In all of
 them we take $\delta_f = 1 \degree$. The values in the color bar have units of
 $day^{-1}$.
  }
 \label{fig:FSLE-gridres}
 \end{figure}

In Figure~\ref{fig:FSLE-gridres} we show an example of the FSLEs
derived using four different F-grids with  increasing resolutions.
 Visually, as the resolution
of the F-grid is increased the structures already observed in the
coarser version are kept, just increasing their detail level. In
addition, when resolution is increased, new less active structures
are appearing in areas previously regarded as almost inactive.
Taking a F-grid finer than the associated V-grid would make
no sense if FSLE was an Eulerian measure obtained from single snapshots
of the velocity field. But FSLE is a Lagrangian measure, i.e. they 
are computed using
trajectories which integrate information on the {\sl history} of the velocity field.
This allows capturing the effects of the large scales on scales smaller
than the V-grid. This does not mean that we reconstruct all the effects
taking place at the smaller scales, but only the ones that have been originated
by the relatively large velocity features which are resolved on the V-grid.


The  main structures of the flow, which
are essentially filaments, become finer as resolution is increased,
behaving much like 
the geometrical persistence of 
a fractal interface 
\cite{Falconer.1990}. The question naturally rises about the possible
multifractal character of FSLEs. Multifractality is a property
characteristic to turbulent flows, and it is associated to the
development of a complex hierarchy accross which energy is transmitted
and utterly dissipated \cite{Frisch.1995,Turiel.2008}.
 The study of
the scaling properties of the distribution of FSLEs at the different
resolution scales reveal that they are multifractal (see Appendix \ref{Ap:Multi}).
 This implies that changes
in scale are accounted for by a well-defined transformation, namely a
cascade multiplicative process \cite{Frisch.1995,Turiel.2008}. It also
implies that information is hierarchized \cite{Turiel.2002} and so
what is obtained first, at the coarsest scales, is the most relevant
information.
Due to multifractality, small-scale structures, as unveiled by the FSLEs,
with typical sizes smaller than that of the velocity resolution, are determined by the
larger ones and the multi-scale invariance properties.
Thus, no artificiality is induced by this calculation and
the robustness of FSLE analysis under changes in scale is confirmed.

\begin{figure}[htb]
 \begin{center}
 \leavevmode
 
 \includegraphics[width=\columnwidth]{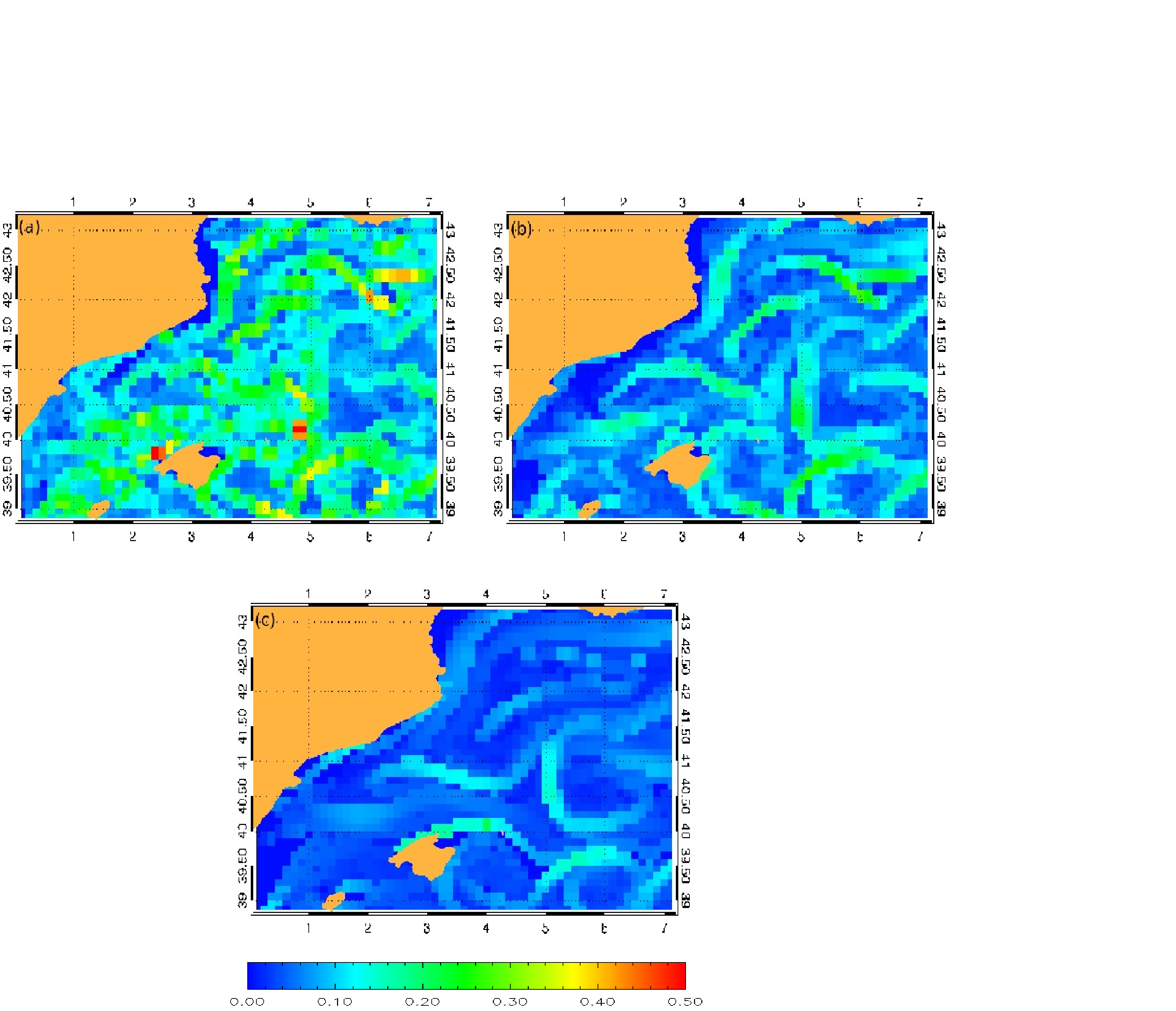}
 \end{center}
 \caption{Snapshots of FSLEs backward in
 time starting from day $640$ at different initial
 V-grid resolutions:
 a)$\Delta_0=1/8 \degree$, b) $\Delta_0=1/4 \degree$,
 c) $\Delta_0=1/2 \degree$. In all of them we take the same F-grid
 resolution of $\delta_0=1/8 \degree$, and $\delta_f =
 1 \degree$. The color bar has units of day$^{-1}$.
  }
 \label{fig:FSLE-velres}
 \end{figure}

\begin{figure}[htb]
\begin{center}
\leavevmode
\includegraphics[width=\columnwidth]{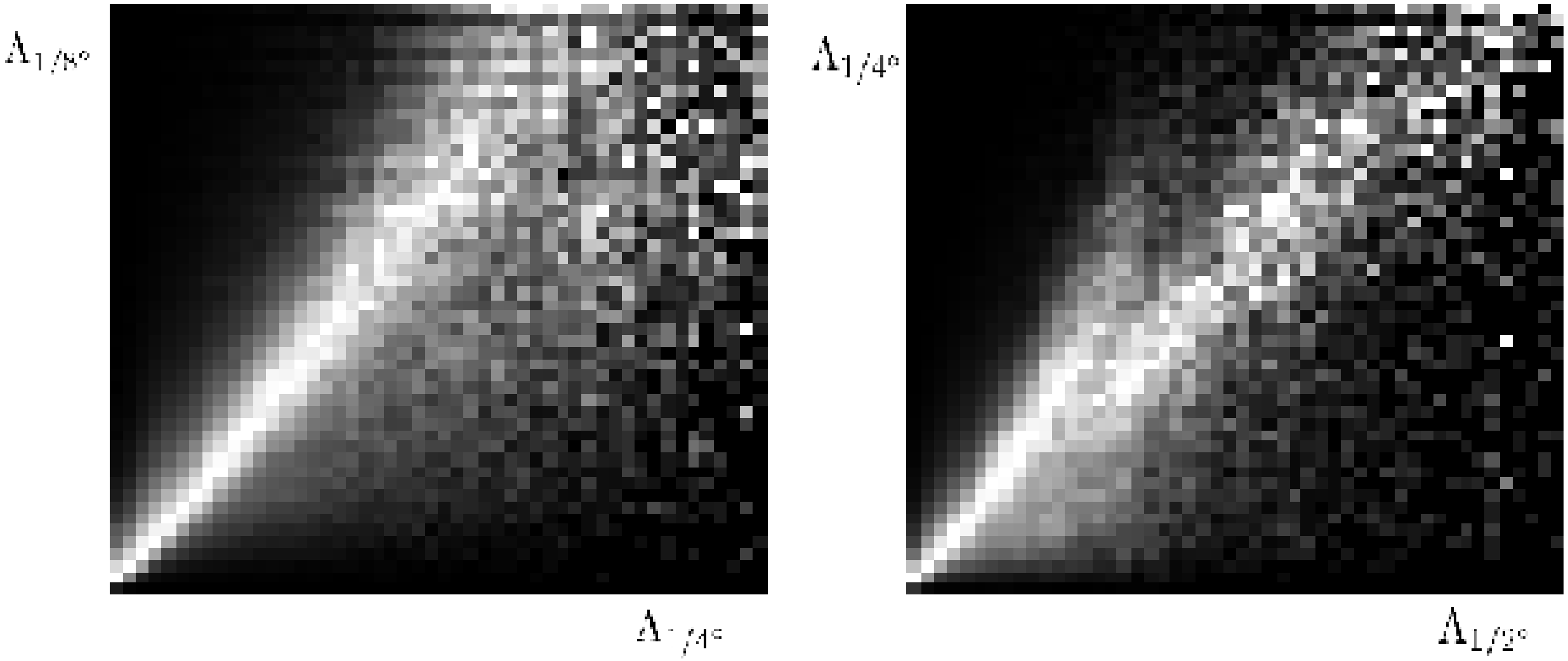}
\end{center}
\caption{Probability distributions (coded as grey levels) of FSLEs derived
at a coarse velocity resolution ($\Lambda_c$, vertical axis) conditioned by
a finer velocity grid ($\Lambda_f$, horizontal axis). The range of values of
both axes run linearly from $0$ to $0.5$ day$^{-1}$. To obtain a statistics
large enough, we have considered $30$ FSLEs snapshots, starting form $t=640$
up to $t=1075$ days, in steps
of $15$ days. The brightest color (pure white) corresponds to the
maximum probability at each column; the darkest color (pure black)
corresponds to zero. {\it Left panel:} FLSEs derived from $1/4
\degree$ velocities conditioned by FSLEs at original $1/8 \degree$
velocities. {\it Right panel:} FSLEs from $1/2 \degree$ velocities
conditioned by FSLEs from $1/4 \degree$ velocities.
}
\label{fig:condFSLE-velres}
\end{figure}

A different question concerns the robustness of FSLEs when the V-grid
is changed. For instance, when a diagnosis is obtained with a
low-resolution velocity field, is this diagnosis compatible
with a later improved observation of the velocity? The answer is
yes. In Figure~\ref{fig:FSLE-velres} we show the FSLEs obtained at a
fixed F-grid resolution of $1/8 \degree$ for varying velocity
resolutions (namely, $1/8\degree$, $1/4\degree$ and $1/2\degree$).
The change of resolution of the velocity grid is performed as indicated
in Appendix \ref{Ap:resolution}.
 We
observe that the global features observed with the coarser resolution
V-grid are kept when this is refined. Obviously,
as the velocity field is described with enhanced resolution new details
(with short-range effect on the flow structure and so no contradicting the
large-scale picture) become apparent in the FSLEs. However, the effect of
refining the V-grid is not only
introducing new small-scale structures: there is a consistent increase
in the values of the FSLEs as the V-grid is refined. The
histograms of the FSLEs, $\Lambda_f$, for a given velocity resolution
conditioned by the FSLEs, $\Lambda_c$, obtained from a coarser velocity field
are shown in  Figure~\ref{fig:condFSLE-velres}. The
modal line (the line of maximum conditioned probability) is close
to a straight line. The best linear regression fits are
$\Lambda_{1/8\degree} =1.08 \Lambda_{1/4 \degree} + 0.05$
(correlation coefficient, $\rho=0.69$) 
and $\Lambda_{1/4\degree} =0.99 \Lambda_{1/2 \degree} + 0.04$
($\rho=0.69$). According to these results, we can approximate the finer
FSLEs $\Lambda_f$ in terms of the coarser FSLEs $\Lambda_c$
as:

\begin{equation}
\Lambda_f (\vec{x})\; =\; \Lambda_c (\vec{x})\: +\: \Delta\Lambda_{fc},
\end{equation}

\noindent
where the quantity $\Delta \Lambda_{fc}$ determines the
contribution to the FSLE by the small-scale variations in velocity not accounted for
by the lower resolution version of this field. It is hence independent of
$\Lambda_c$, so the intercept of the vertical axis with the linear regression equals the mean
of this quantity.

%

A linear dependence of $\Lambda_f$ with $\Lambda_c$ when scale
is changed implies that FSLEs
follow a multiplicative cascade \cite{Turiel.2006,Turiel.2008}, an
essential ingredient in multifractal systems which gives further confirmation to
our previous results.
We have hence shown that
a) the dependence of FSLEs on both types of scale parameters reveals a
multifractal structure; b) what is diagnosed at the coarser scales is
still valid when scale is refined (although as V-grid resolution is
increased the reference level of FSLEs is increased by a constant).

\section{Effect of noise on FSLEs}
\label{Sec:noise}

\begin{figure}[htb]
 \begin{center}
 \leavevmode
 \includegraphics[width=7.5cm]{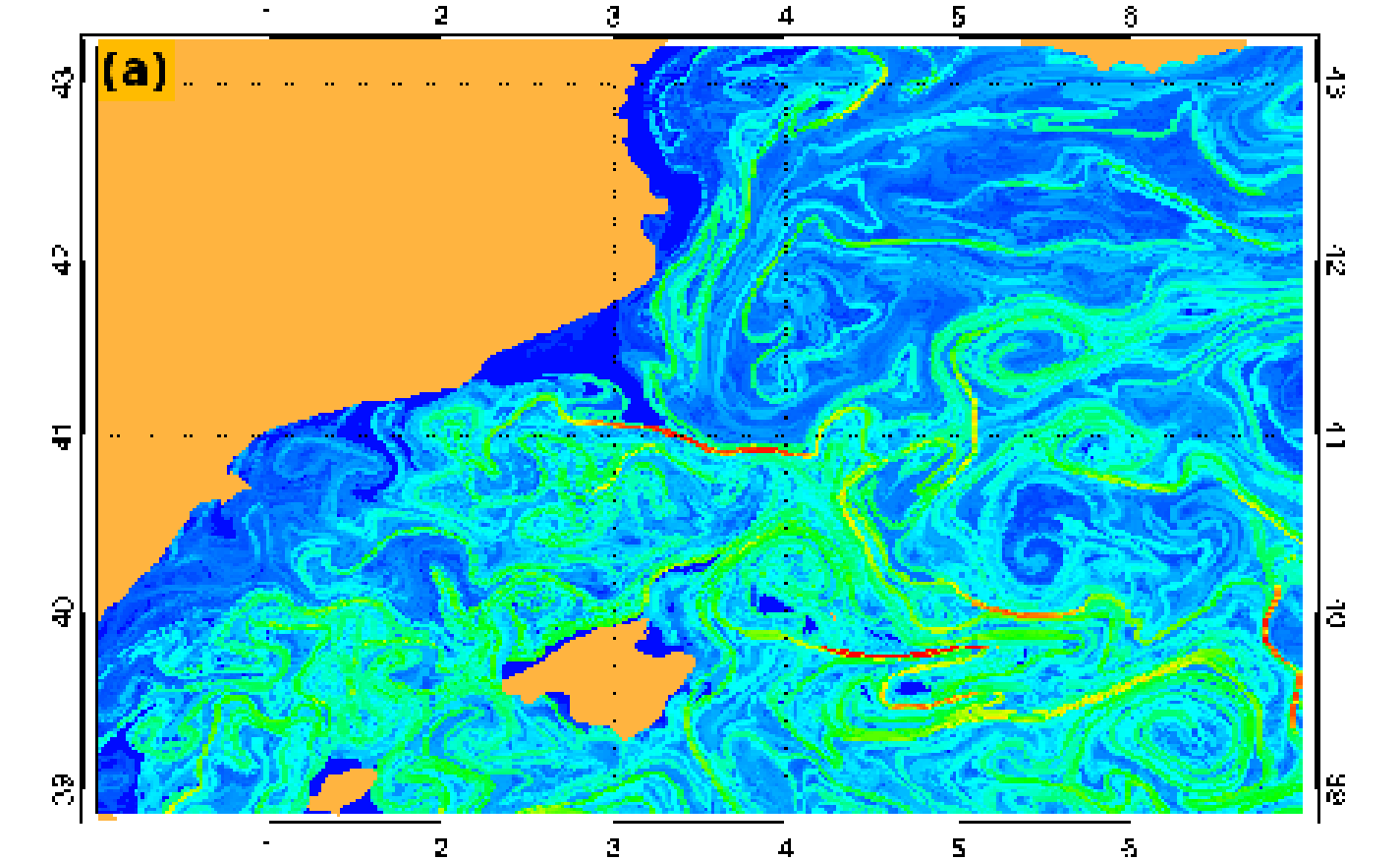}
 \includegraphics[width=7.5cm]{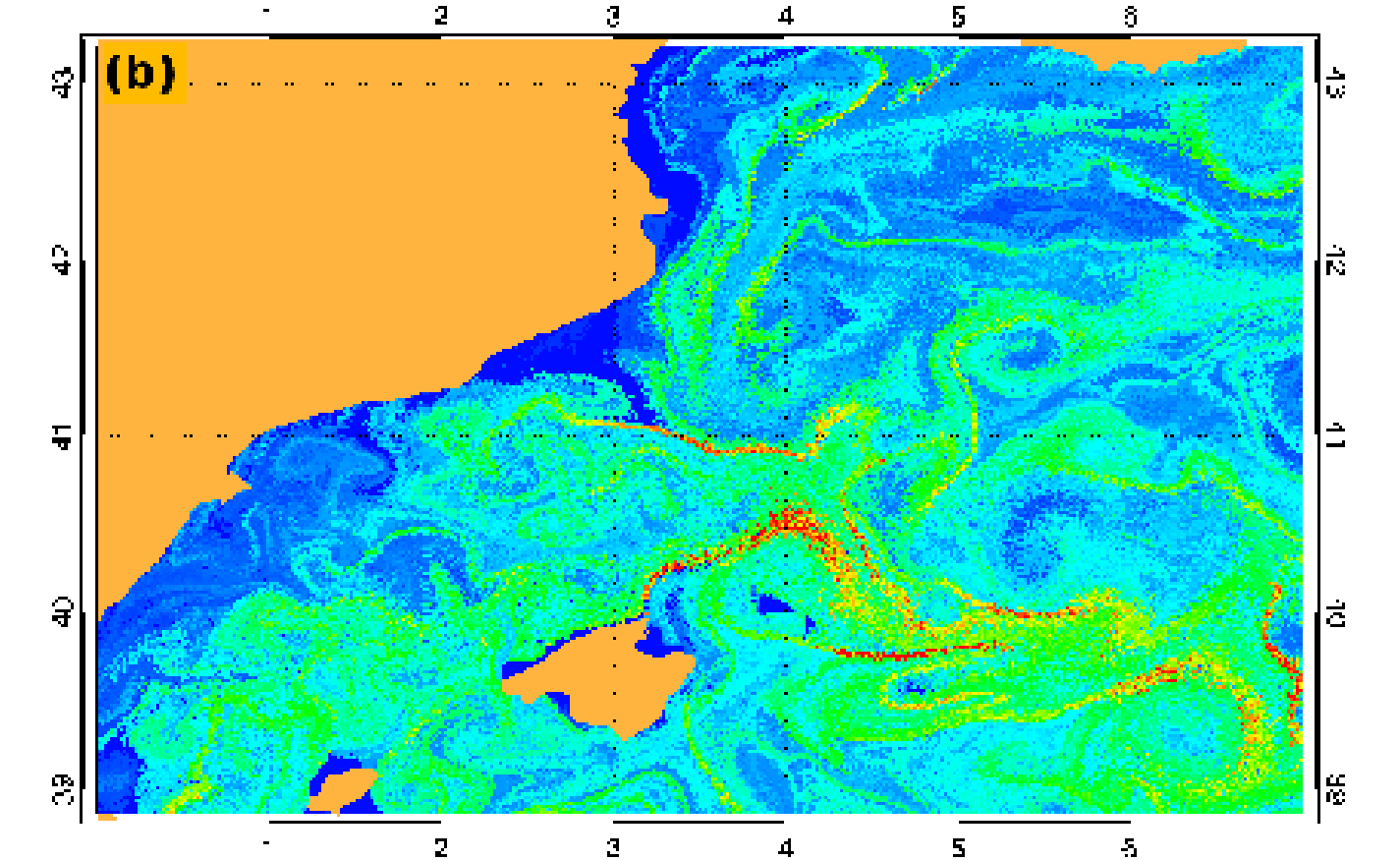}
 \\
 \leavevmode
 \includegraphics[width=10cm,height=1cm]{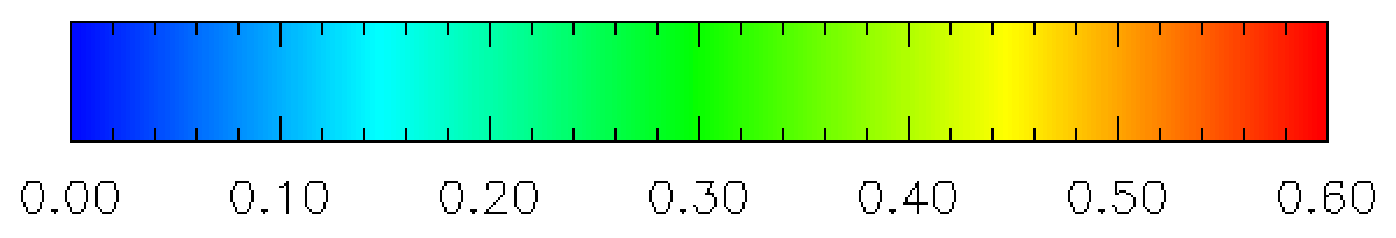}
 \end{center}
 \caption{Snapshots of FSLEs calculated backwards in time starting from day $600$
 at fixed spatial resolution
 ($\delta_0=1/64 \degree$), and at different $\alpha$:
  a)$\alpha = 0$, b) $\alpha = 10$.
 In both of them we take
 $\delta_f = 1 \degree$. The color bar has units of $day^{-1}$.
 Initial conditions for which the separation $\delta_f$ has not
 been reached after $600$ days are assigned a value $\Lambda=0$.
  }
 \label{fig:errorsnapshot}
 \end{figure}

We compute the FSLEs after applying a random perturbation to all
components of the velocity field. 
The velocity is changed from
$(u,v)$ to $(u',v')$, with $u'(x,t)=
u(\textbf{x},t)(1+\alpha\eta_x(\textbf{x},t))$ and $v'(\textbf{x},t)=
v(\textbf{x},t)(1+\alpha\eta_y(\textbf{x},t))$.
$\{\eta_x(\textbf{x},t),\eta_y(\textbf{x},t)\}$ are sets of Gaussian random
numbers of zero mean and unit variance. $\alpha$ measures the
relative size of the perturbation
(it 
gives the ratio of the mean
amplitude of noise with respect to mean amplitude of the velocity).
 We introduce three different
kinds of error: uncorrelated noise, i.e. different and
uncorrelated values of $\{\eta_x(\textbf{x},t),\eta_y(\textbf{x},t)\}$ for each
$\textbf{x}$ and $t$; correlated in time and uncorrelated in space
(uncorrelated for different $\textbf{x}$ but the same values at given
$\textbf{x}$ for different $t$); and correlated in space and uncorrelated
in time (uncorrelated values for different $t$, but the same
values for different $\textbf{x}$ at fixed $t$). Note that the
perturbation is proportional to the original velocity.
Figure \ref{fig:errorsnapshot} shows a snapshot of FSLEs
for the velocity field 
 perturbed by
uncorrelated noise of a relative size $\alpha=10$, i.e., noise is $10$ times
larger than the amplitude of the initial velocity field.
The computed Lagrangian structures look rather the same, despite the large size
of the perturbation introduced.

%

\begin{figure}[htb]
\begin{center}
\leavevmode
\includegraphics[width=6cm,angle=270]{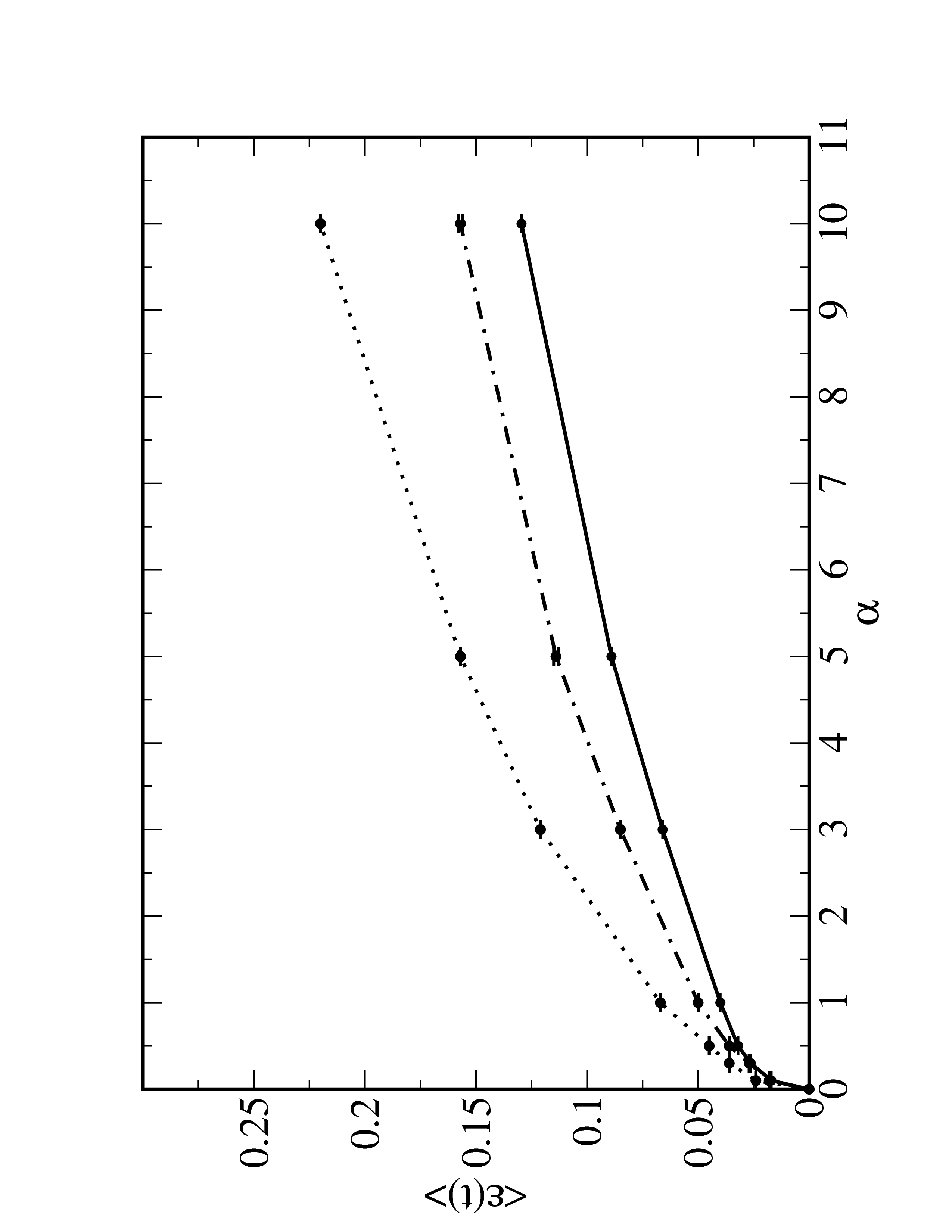}
\includegraphics[width=6cm,angle=270]{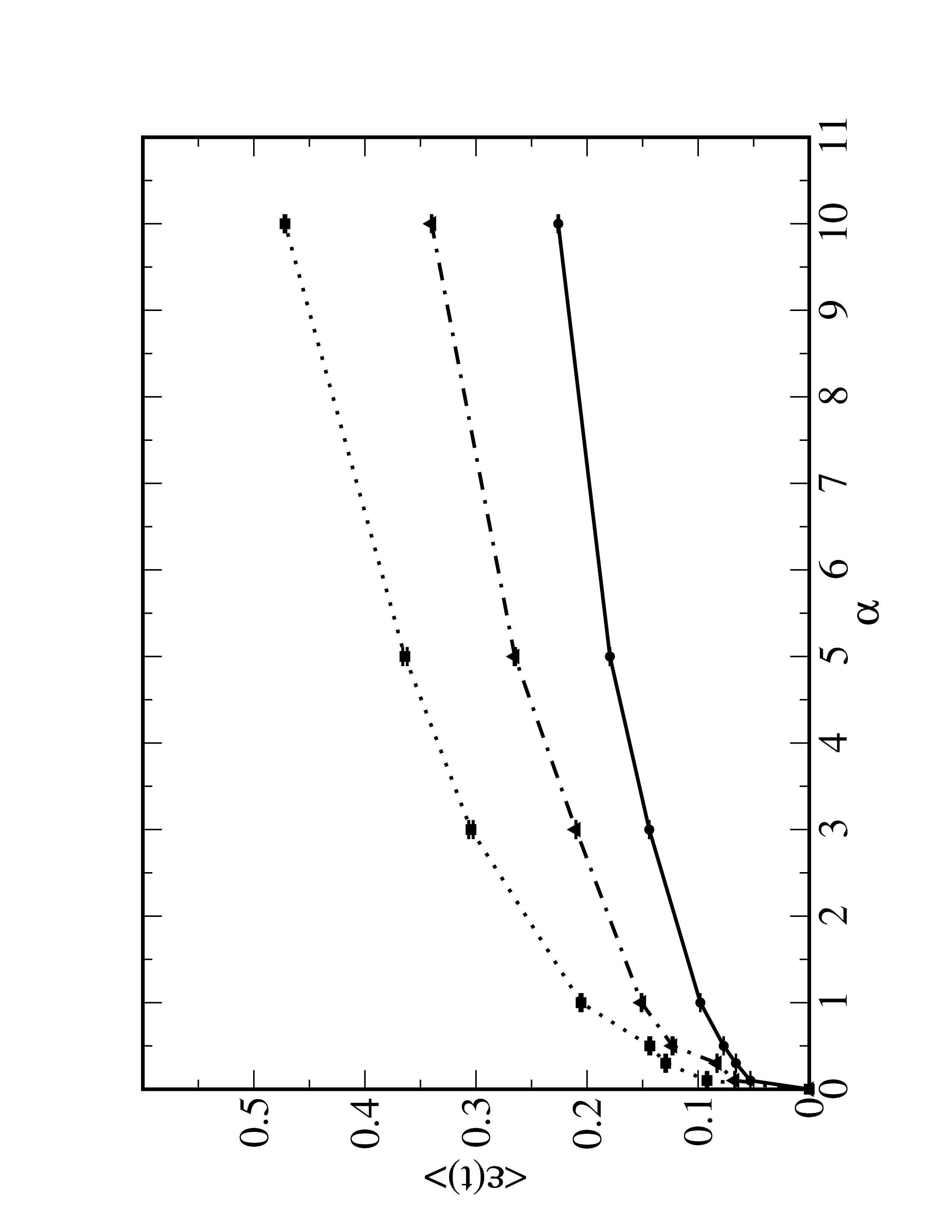}
\end{center}
\caption{Relative error $<\epsilon(t)>$ of the FSLE fields for different
perturbation intensity $\alpha$ in the velocity data. Solid line
is for uncorrelated noise in space and time, dashed-dotted line is
for uncorrelated noise in time and correlated in space, and dotted
line is for uncorrelated noise in space and correlated in time.
$<\epsilon(t)>$ is obtained by averaging the RE in $100$
snapshots (see Eq. (\ref{eq:error})). The error bar is the
statistical error of the temporal average $<\epsilon(t)>$. Left:
spatial resolution $\delta_0=1/8 \degree$. Right: spatial
resolution $\delta_0=1/64 \degree$. In
all calculations we take $\delta_f = 1 \degree$
 }
\label{fig:error}
\end{figure}

In order to quantify the influence
of the velocity perturbation in the FSLE
calculation we compute the relative error (RE) of perturbed FSLEs with
respect to unperturbed FSLEs,  $<\epsilon(t)>$, 
 at a given instant of
time, and then averaging in time (we have $M=100$ snapshots $:
t=t_1,...,t_M$) as:
\begin{equation}
\epsilon(t_i)=\sqrt{\frac{1}{N} \sum_{\textbf{x}} \dfrac{|\Lambda^\alpha
(\textbf{x},t_i) - \Lambda (\textbf{x},t_i)|^2}{|\Lambda(\textbf{x},t_i)|^2}}, \ \ \
\hfill <\epsilon(t)> \equiv \frac{1}{M} \sum_{i=1}^M \epsilon(t_i)
\ .
\label{eq:error}
\end{equation}
$\Lambda(\textbf{x},t_i)$ and $\Lambda^\alpha(\textbf{x},t_i)$ are the FSLEs
fields without and with inclusion of the perturbation in the
velocity data, respectively. The sum over $\textbf{x}$ runs over the
$N=2679$ spatial points.
Figure~\ref{fig:error} shows the average RE as a function of
$\alpha$. It must be remarked that the RE has always small
values: even for $\alpha=10$  the RE remains
smaller than $0.23$ for the three kinds of noise.
To get an idea of how  relevant these quantities are,
we have computed the RE of shuffled FSLEs (permuting locations
at random) with respect to the original ones, and obtained a value of
$1.143$. FSLEs are thus robust against uncorrelated noise; the reason
is the averaging effect produced when computing them
by integrating over trajectories which extend in time and space, that tends to
cancellate random, uncorrelated errors.

Now we proceed by adding noise to the particle trajectories.
This is a simplified way of including unresolved small scales in
the Lagrangian computations \cite{Griffa.1996}. To be precise we
solve numerically the system of Equations~(\ref{eq:noise1}) and
(\ref{eq:noise2}) (see Appendix \ref{Ap:noise}), where a stochastic term  with a
Gaussian random number and an effective eddy-diffusion,  $D$, has been added.
For the diffusivity we use Okubo's empirical formula \cite{Okubo.1971},
which relates the effective eddy-diffusion, $D$ in
$m^2/s$, with the spatial scale, $l$ in meters: $D(l)= 2.055 \
10^{-4} \ l^{1.15}$. If we take $l=12 \ km$, which is the
approximate length corresponding to the $1/8 \degree$ DieCAST
resolution at Mediterranean latitudes, we obtain $D \sim 10 \ m^2
s^{-1} \equiv D_0$.

\begin{figure}[htb]
\begin{center}
\leavevmode
\includegraphics[width=12cm]{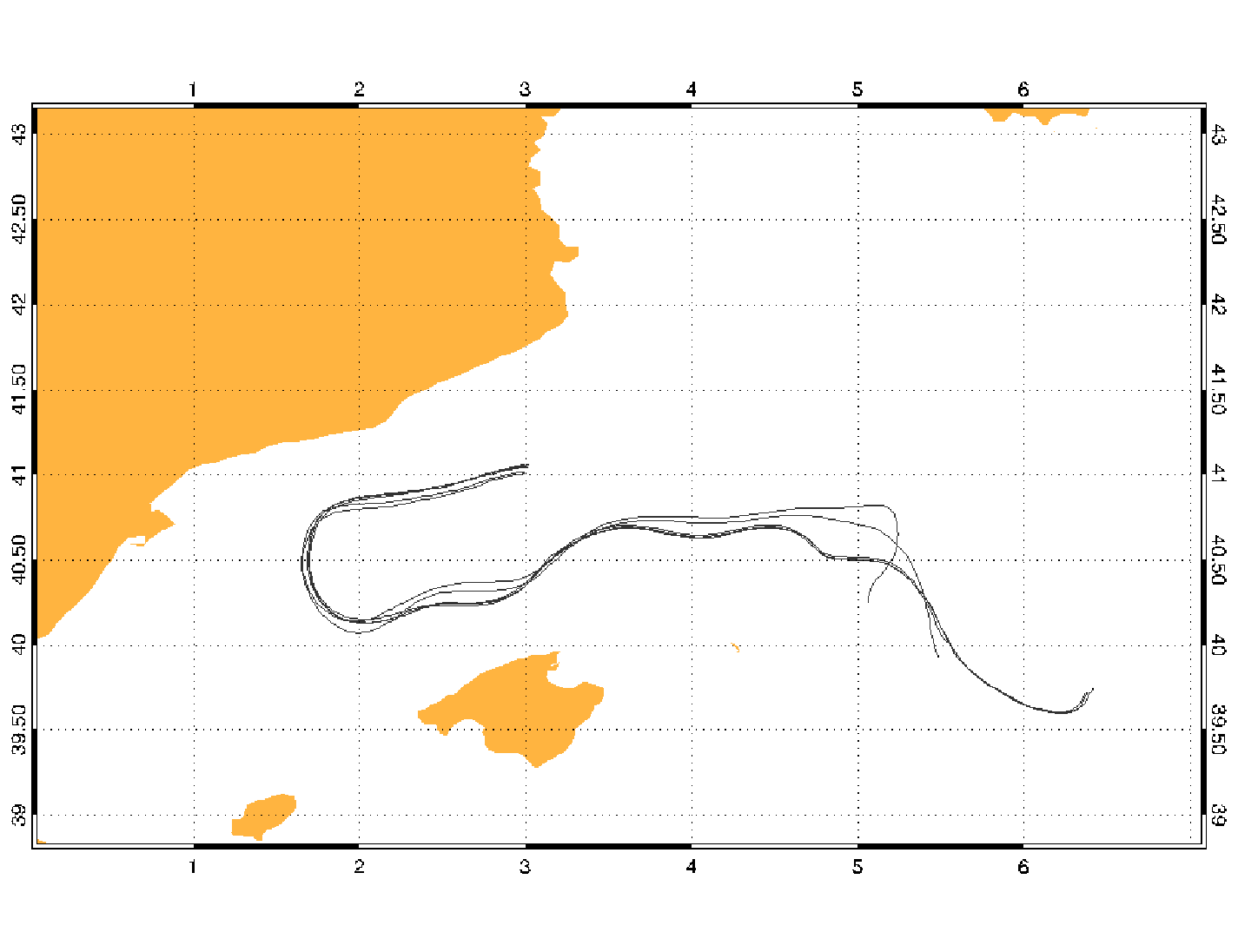}
\\
\leavevmode
\includegraphics[width=12cm]{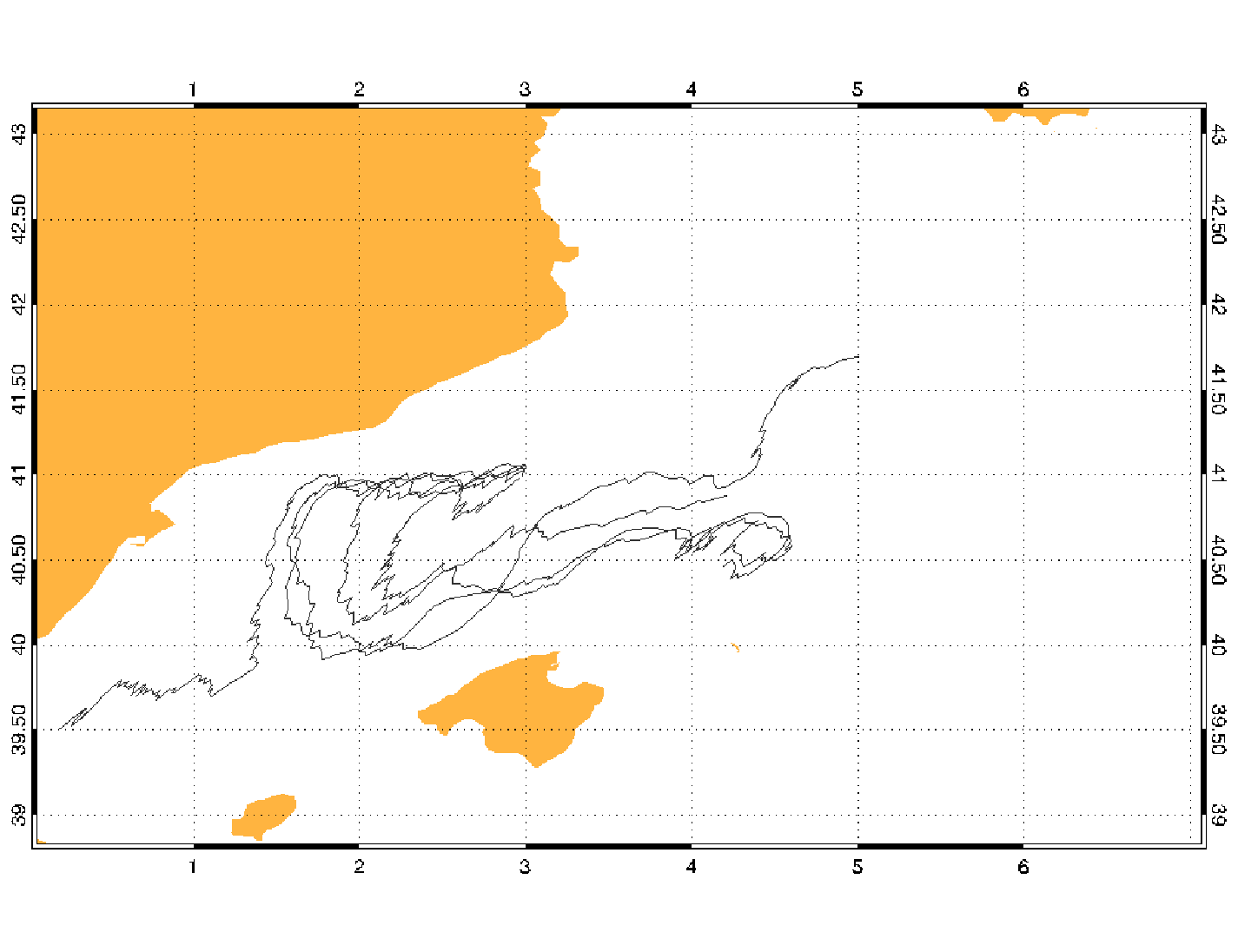}
\end{center}
\caption{Trajectories of five particles without diffusion (top)
and with diffusion (bottom). The difference in the initial positions
of all five particles is about $0.06 \degree$, and we use these initial conditions
in both computations. The
trajectories were computed for $50$ days of integration. We used the
eddy-diffusion $D_0 \sim 10 m^2 s^{-1}$ assigned by the Okubo
formula to the resolution of the DieCAST model at Mediterranean
latitudes.
 }
\label{fig:traj}
\end{figure}

\begin{figure}[htb]
 \begin{center}
 \leavevmode
 \includegraphics[width=7.5cm]{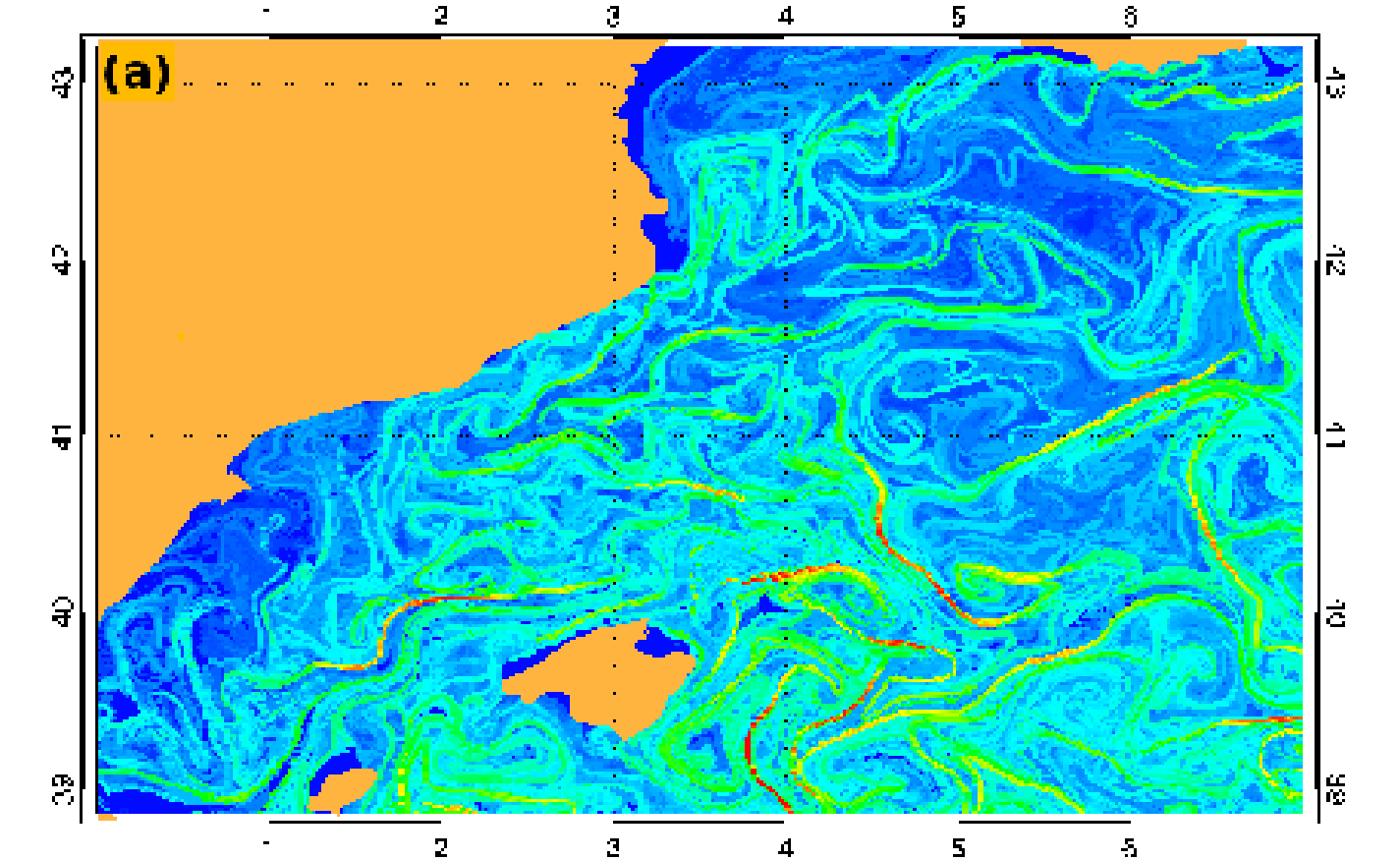}
 \includegraphics[width=7.5cm]{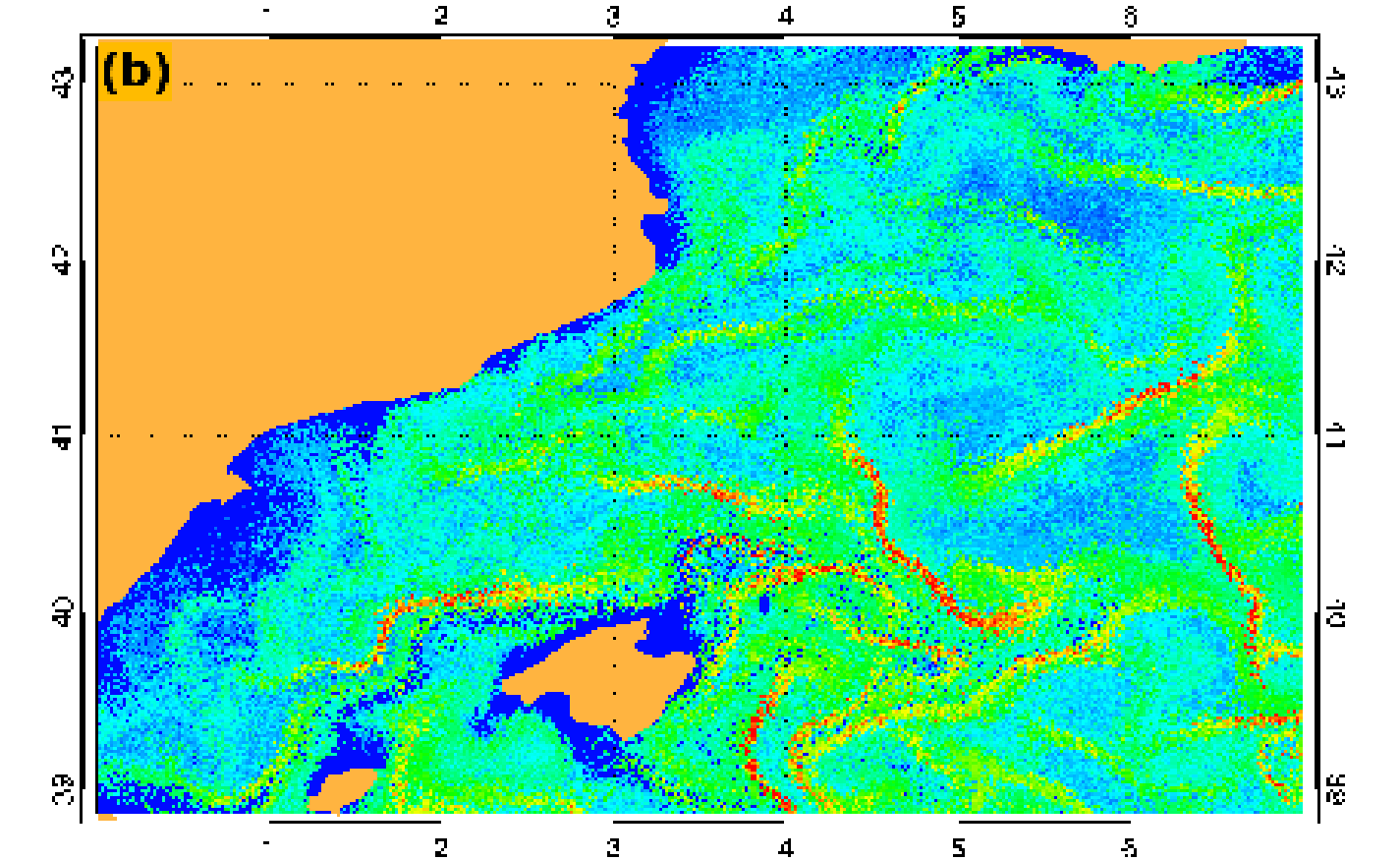}
 \\
 \leavevmode
 \includegraphics[width=10cm,height=1cm]{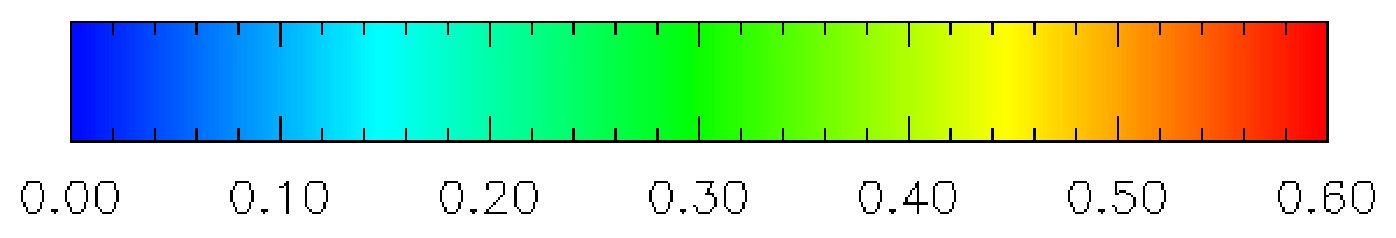}
 \end{center}
 \caption{FSLEs computed backwards from day $500$
  at the same spatial resolution
 ($\delta_0=1/64 \degree$), and for different eddy-diffusion
 values: a)$D = 0 ~m^2 s^{-1}$ b) $D = 0.9 ~m^2 s^{-1}$.
 We take
 $\delta_f = 1 \degree$. The color bar has units of $day^{-1}$.
 Initial conditions for which the separation $\delta_f$ has not
 been reached after $500$ days are assigned a value $\Lambda=0$.
  }
 \label{fig:diffusion}
 \end{figure}

\begin{figure}[htb]
\begin{center}
\leavevmode
\includegraphics[width=16cm]{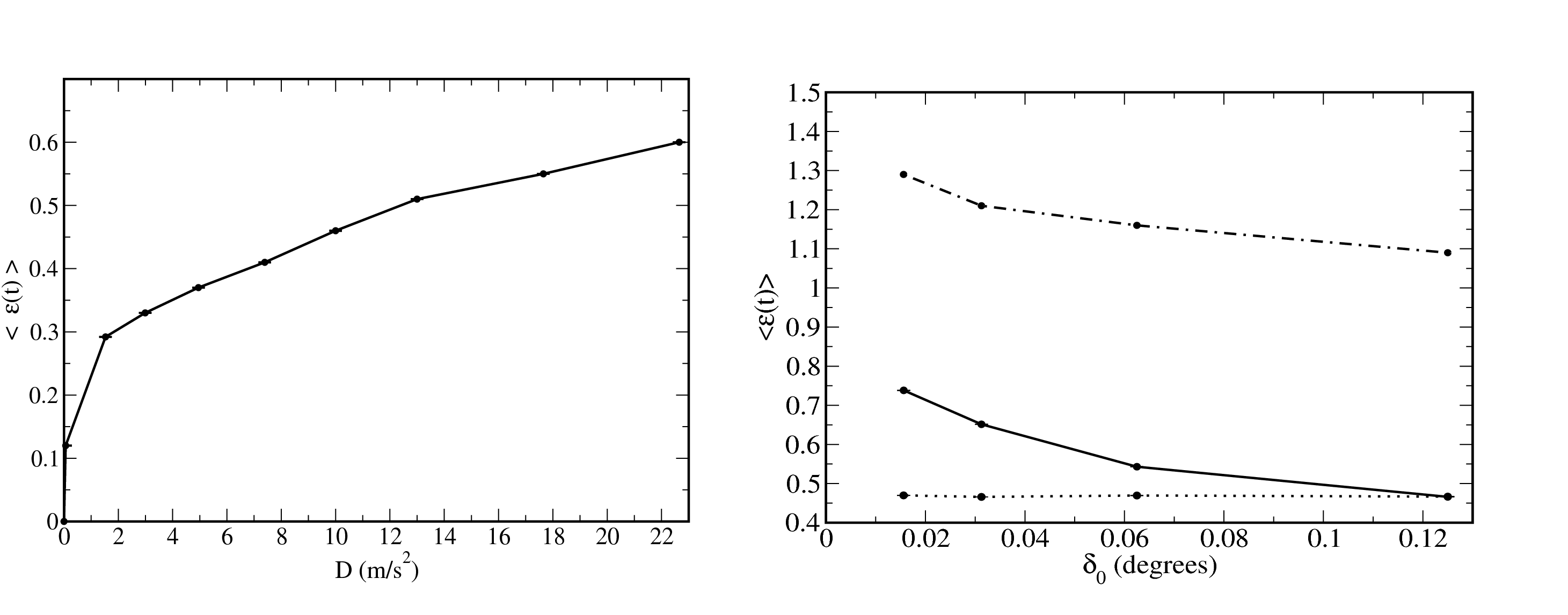}
\end{center}
\caption{Left: Relative
error $<\epsilon(t)>$ of the FSLE at the different values of $D$
in the particle trajectories, with respect to the $D=0$ case.
Spatial resolution is $\delta_0=1/8 \degree$, and $\delta_f = 1
\degree$. $<\epsilon(t)>$ is obtained by temporally averaging the
relative errors in $100$ snapshots. The (small) error bars indicate
the statistical error in the $<\epsilon(t)>$ average.
Right: Dotted line is the RE
 $<\epsilon(t)>$ of the FSLE at
different spatial resolution $\delta_0$ and at one assigned
eddy-diffusion $D$ for every spatial resolution in the particle
trajectories with respect to the $D=0$ case. Solid line is the RE $<\epsilon(t)>$ of
the FSLE at different spatial resolution $\delta_0$, and at the
same eddy-diffusion $D_0 = 10 m^2 s^{-1}$ in the particle
trajectories with respect to the $D=0$ case. Dashed-dotted line is the RE of shuffled
FSLE with respect to the original case ($D=0$) at different spatial resolution. $<\epsilon(t)>$ by
temporally averaging the RE in $100$ snapshots. The (small)
error bar indicates the statistical error in the $<\epsilon(t)>$
average. In all of them we take $\delta_f = 1 \degree$.
}
\label{fig:errordifus}
\end{figure}

In Figure~\ref{fig:traj} we show particle trajectories
without (top panel) and with (bottom panel) eddy diffusion.
As expected diffusion introduces small scale irregularities on the
trajectories, and also substantial dispersion at large scales.
In Figure~\ref{fig:diffusion},  FSLEs with $\delta_0=1/64 \degree$
$D = 0 ~m^2 s^{-1}$ and $D = 0.9 ~m^2 s^{-1}$
 (obtained for this scale by
Okubo's formula for spatial resolution) are shown. We can see that
the main mesoscale structures are maintained, but small-scale filamental
structures are lost since filaments become blurred. This is somehow expected
a priory because diffusion introduces a new length scale $l_D$ proportional to
$\sqrt{D}$. A pointwise comparison of noiseless and noise-affected
FSLEs makes no sense, since the noise-induced blurring disperses FSLEs values
specially at places with low values, see
Figure~\ref{fig:diffusion}. But for Lagrangian diagnostics high FSLE values
are much more relevant, so we compute the error restricted to the places
where $\Lambda>0.2$ for $D=0$. Left panel of 
Figure~\ref{fig:errordifus} shows the RE respect to the $D=0$ case
(applying Equation~(\ref{eq:error})) for
different values of $D$. The RE monotonously increases
with $D$, but remains smaller than $0.6$ for the largest value of
$D$ considered. Again, to get an idea of how relevant these relative
errors are, we have to compare them with the RE of shuffled FSLEs, of
value
$1.143$.

%

As a matter of fact, diffusion introduces an effective observation scale, and
one should not go beyond that limit to obtain senseful results; this is
illustrated in the right panel of Figure~\ref{fig:errordifus}.
As shown in the figure, for fixed eddy diffusivity  (in the case of the
figure, $D_0 = 10 m^2 s^{-1}$), when $\delta_0$ becomes
greater the error diminishes. Hence, a fixed diffusion will eventually be
negligible at a scale large enough, in this way determining an observation scale. A
completely different situation is given when the eddy-diffusion depends
on the scale according to Okubo's formula (in our case,
at $\delta_0=1/16 \degree$ $D = 4,5 m^2 s^{-1}$; at $\delta_0=1/32
\degree$ $D = 2 m^2 s^{-1}$; and at $\delta_0=1/64 \degree$, $D = 0.9
m^2 s^{-1}$). Now $<\epsilon(t)>$ takes a constant value close to
$0.45$, meaning that Okubo's diffusion behaves the same at all scales.
 This is expected since Okubo's law is based on the hypothesis that unsolved scales act
as  diffusers, like in turbulence \cite{Frisch.1995}.
Our result is consistent with the ideas behind Okubo's hypothesis.

\begin{figure}[htb]
\begin{center}
\leavevmode
\includegraphics[width=14cm,angle=270]{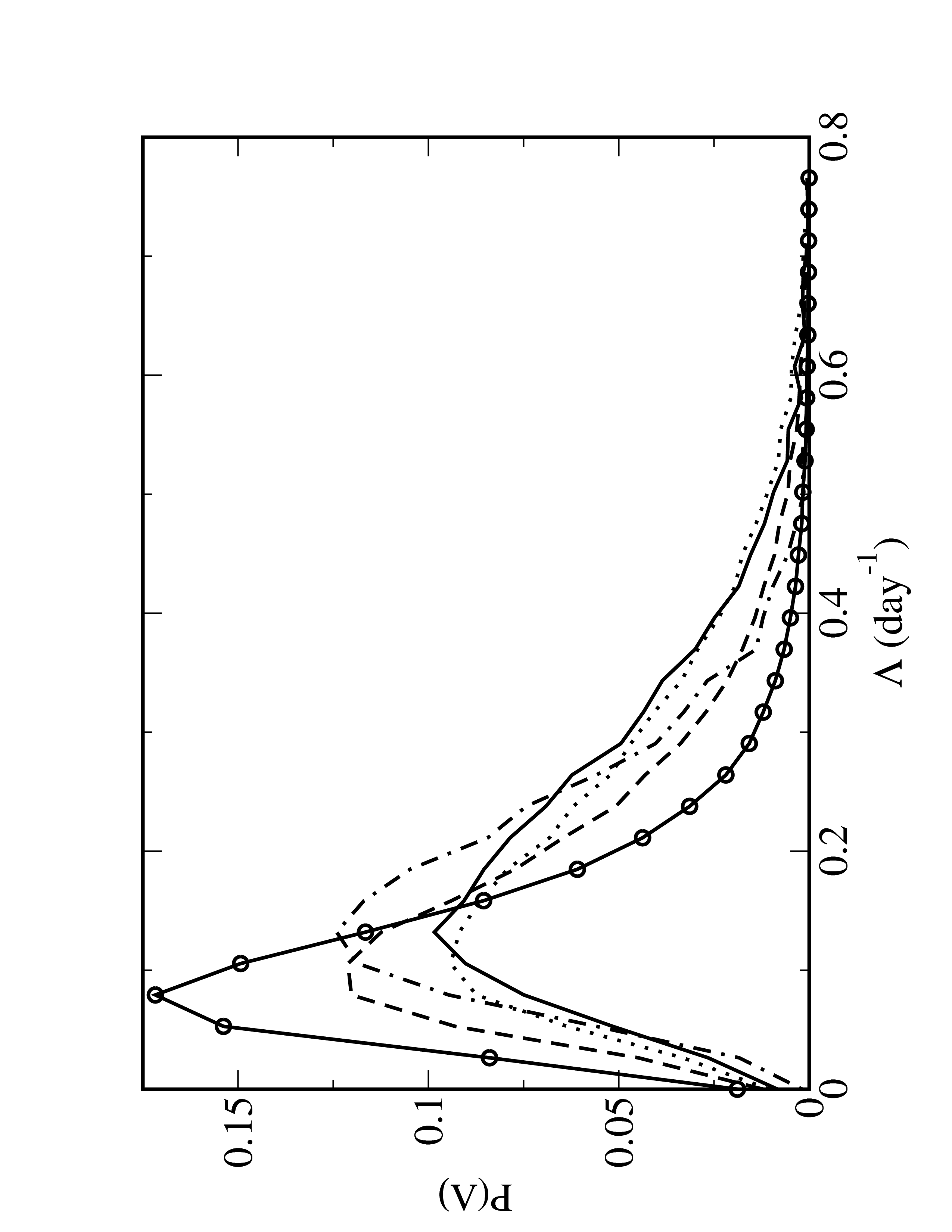}
\end{center}
\caption{Comparison between probability density function for the
FSLEs at different resolutions with different values of
eddy-diffusion, and without diffusion. It is obtained from the
temporal average ($30$ snapshots) of histograms. Solid line for
$\delta_0=1/8 \degree $ with diffusion, dotted $\delta_0=1/8
\degree$ without diffusion, dashed $\delta_0=1/16 \degree$ without
diffusion, dashed-dotted line for $\delta_0=1/64 \degree$ with
diffusion, and circle-line for $\delta_0=1/64 \degree$ without
diffusion. }
\label{fig:histodifusion}
\end{figure}

We can characterize the effective diffusion scale from the properties of
the histograms.
In Figure~\ref{fig:histodifusion}
we present the histograms of FSLEs at different F-grid resolutions, and including
or not eddy diffusion which takes always the same value $D_0 = 10 m^2 s^{-1}$.
For $\delta_0 =1/8 \degree$ the histograms with and without diffusion are almost
coincident. This is due to the fact that the value of diffusion we are using
is the one corresponding, by the Okubo's formula, to $1/8 \degree$. i.e., we are
parameterizing turbulence below $1/8 \degree$, and this has no
effects on the FSLE computations if the minimum scale considered
is also $1/8$. However, this behavior is different for smaller $\delta_0$
(always keeping $D_0=10m^2 s^{-1}$). The histograms for
$\delta_0=1/16, 1/64 \degree$, with and without diffusion,
are clearly different, those including diffusion becoming closer to the
histogram for $\delta_0=1/8 \degree$.

\section{Conclusions}
\label{sec:conclusions}

In this paper, we have analyzed the sensibility of FSLE-based analyses for the
diagnostic of Lagrangian properties of the ocean (most notably, horizontal
mixing and dispersion). Our sensibility tests include the two most important
effects when facing real data, namely dynamics of unsolved scales and of noise.
Our results show that even if some dynamics are missed (because of lack
of sampling or inaccuracy of any kind in the measurements) FLSEs results would
still give an accurate picture of Lagrangian properties, valid for the solved
scales. This does not mean that scale and/or noise leave FSLEs unaffected, but
the way in which they modify this Lagrangian diagnostics can be properly
accounted.

Lagrangian methods provide answers to problems which have a deep impact
on risk management (e.g. control of pollutant dispersion)  as well as
on ecosystem analysis (e.g. tracking nutrient mixing and transport,
identifying the role of horizontal mixing in primary
productivity). They utterly will give hints about energy exchanges in
the upper ocean and will help in  understanding processes driving
global change in the oceans. The use of Lagrangian techniques for the
assessment of the transport and mixing properties of the ocean has
grown in importance in the latest years, with increasing efforts devoted to
the implementation of appropriate techniques but few
studies on the validity of the results when real data, affected by
realistic constraints, are used. Our work will serve to unify and interpret the
analyses provided by Lagrangian methods when real data are processed.

\acknowledgments

This work is a contribution to OCEANTECH project (CSIC PIF-2006).
IH-C, CL and EH-G acknowledge support from project
FISICOS (FIS2007-60327) of MEC and FEDER.

\appendix

\section{Calculation of FSLE.}
\label{Ap:FSLE}

It is natural to choose the initial points $(x,y)$ on the nodes of a
grid with lattice spacing coincident with the initial separation of
fluid particles $\delta_0$. In this way one obtains maps of values of
$\Lambda$ at a spatial resolution that will coincide with
$\delta_0$. To compute $\Lambda$ we need to know the trajectories of
the particles. The equations of motion that describe the horizontal
evolution of particle trajectories in the velocity field are:

\begin{eqnarray}
\frac{d\phi}{dt}&=&\frac{u(\phi, \lambda, t)}{R \cos {\lambda}},
\label{eq:motion1}\\
\frac{d\lambda}{dt}&=&\frac{v(\phi, \lambda, t)}{R},
\label{eq:motion2}
\end{eqnarray}

\noindent
where $u$ and $v$ represent the zonal and meridional components of the
surface velocity field coming from the simulations described above,
$R$ is the radius of the Earth (6400 km in our
computations). Numerically we proceed integrating Eqs.~(\ref{eq:motion1})
and (\ref{eq:motion2}) using a
standard fourth-order Runge-Kutta scheme with an integration time step
$dt$ = 6 hours. Spatiotemporal interpolation of the velocity data is
achieved by bilinear interpolation. Since this technique requires an
equally spaced grid and this is not the case for the spherical
coordinates $(\phi,\lambda)$, for which
the grid is not uniformly spaced in the latitude coordinate, we first
transform it to a new system $(\phi,\mu)$ where the grid turns out to
be uniformly sampled \cite{Mancho.2008}. The latitude $\lambda$ is
related to the new coordinate $\mu$ according to:

\begin{equation}
\mu\; =\; \log |\sec \lambda \: +\: \tan\lambda|
\label{eq:mu}
\end{equation}

Using the new coordinate variables the equations of motion become:

\begin{eqnarray}
\frac{d\phi}{dt}&=&\frac{u(\phi, \mu, t)}{R \cos {\lambda(\mu)}}
\label{eq:motion-simplified1}
\\
\frac{d\mu}{dt}&=&\frac{v(\phi, \mu, t)}{R\cos {\lambda(\mu)}},
\label{eq:motion-simplified2}
\end{eqnarray}

\noindent
and one can convert the $\mu$ values back to latitude $\lambda$ by
inverting Eq.~(\ref{eq:mu}): $\lambda = \pi/2-2 \arctan e^{-\mu}$. Once we
integrate the equations of motion, Eqs.~(\ref{eq:motion-simplified1}) and
(\ref{eq:motion-simplified2}), we compute the FSLEs
applying Eq.~(\ref{eq:FSLE}) for the points of a lattice with spacing
$\delta_0$. We will only compute FSLEs
integrating backwards-in-time the particle trajectories, since
LCSs associated to this has a direct physical interpretation, but
all our results are similar for forward-in-time dynamics.

\section{Multifractal character of FSLE.}
\label{Ap:Multi}

To study if FSLEs behave like multifractal systems we have
computed their probability density distribution, 
$P(\delta_0, \Lambda)$, for different resolutions $\delta_0$.
It must follow that for any resolution scale $\delta_0$ of
FSLE grid we should observe:

\begin{equation}
P(\delta_0,\Lambda) \propto \delta_0^{d-D(\Lambda)},
\label{eq:FSLEpdf}
\end{equation}

Notice that as $d-D(\Lambda)$ is a positive quantity, 
$P (\delta_0, \Lambda)$
 becomes
smaller as $\delta_0$ is reduced. In fact, a characteristic signature
of multifractal scaling is having a scale-dependent histogram which
becomes more strongly peaked as the resolution scale becomes smaller.
We know that there is a FSLE for each domain point, so we can
normalize the FSLE distribution by its maximum (that will be attained
for a value $\Lambda_c$) and then retrieve the
associated singularity spectrum according to the following expression:

\begin{equation}
D(\Lambda) \; =\; d-\frac{\log{\frac{P(\delta_0, \Lambda)}{P(\delta_0,
      \Lambda_c)}}}
{\log{\delta_0}}.
\label{eq:Dlambda}
\end{equation}

\begin{figure}[htb]
\begin{center}
\leavevmode
\includegraphics[width=12cm]{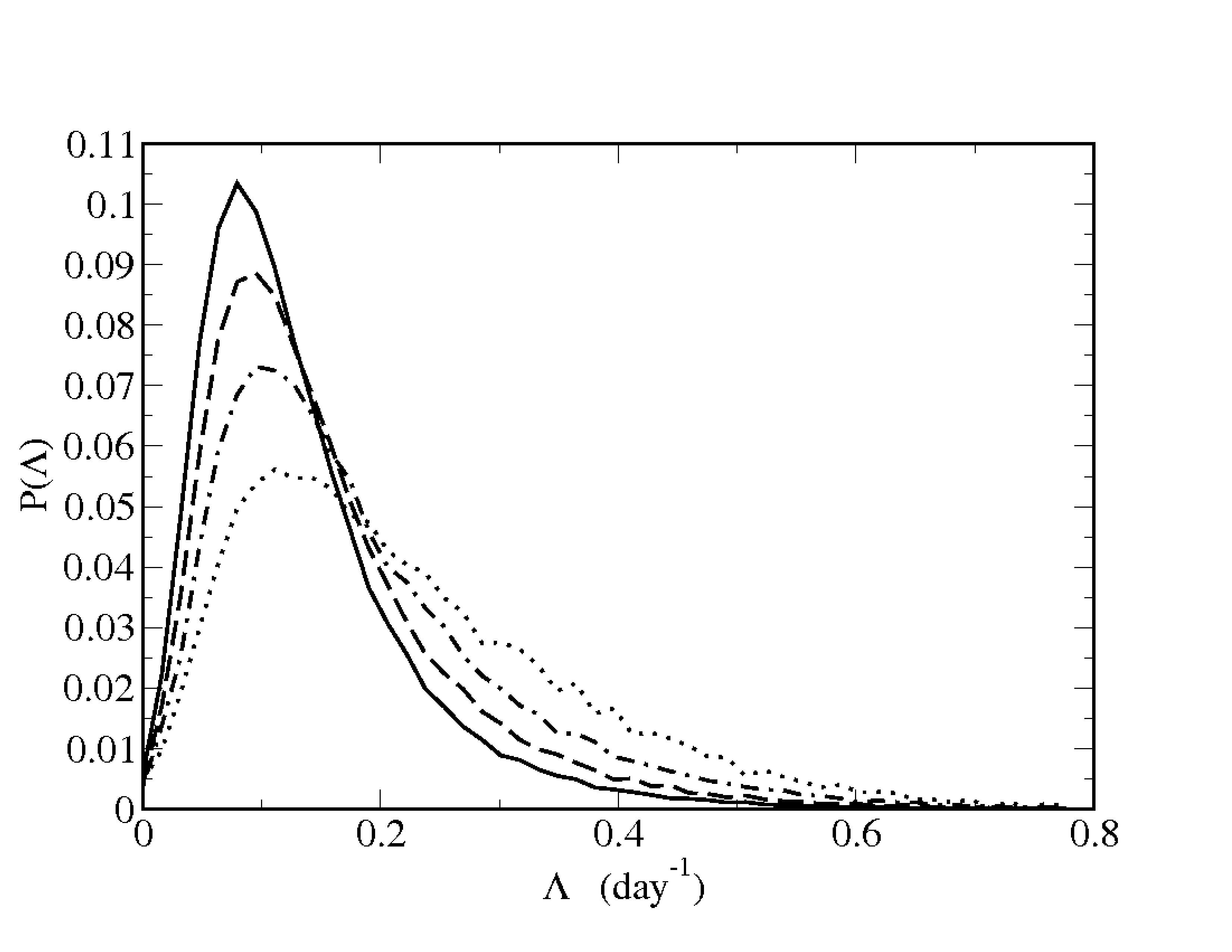}
\\
\leavevmode
\includegraphics[width=12cm]{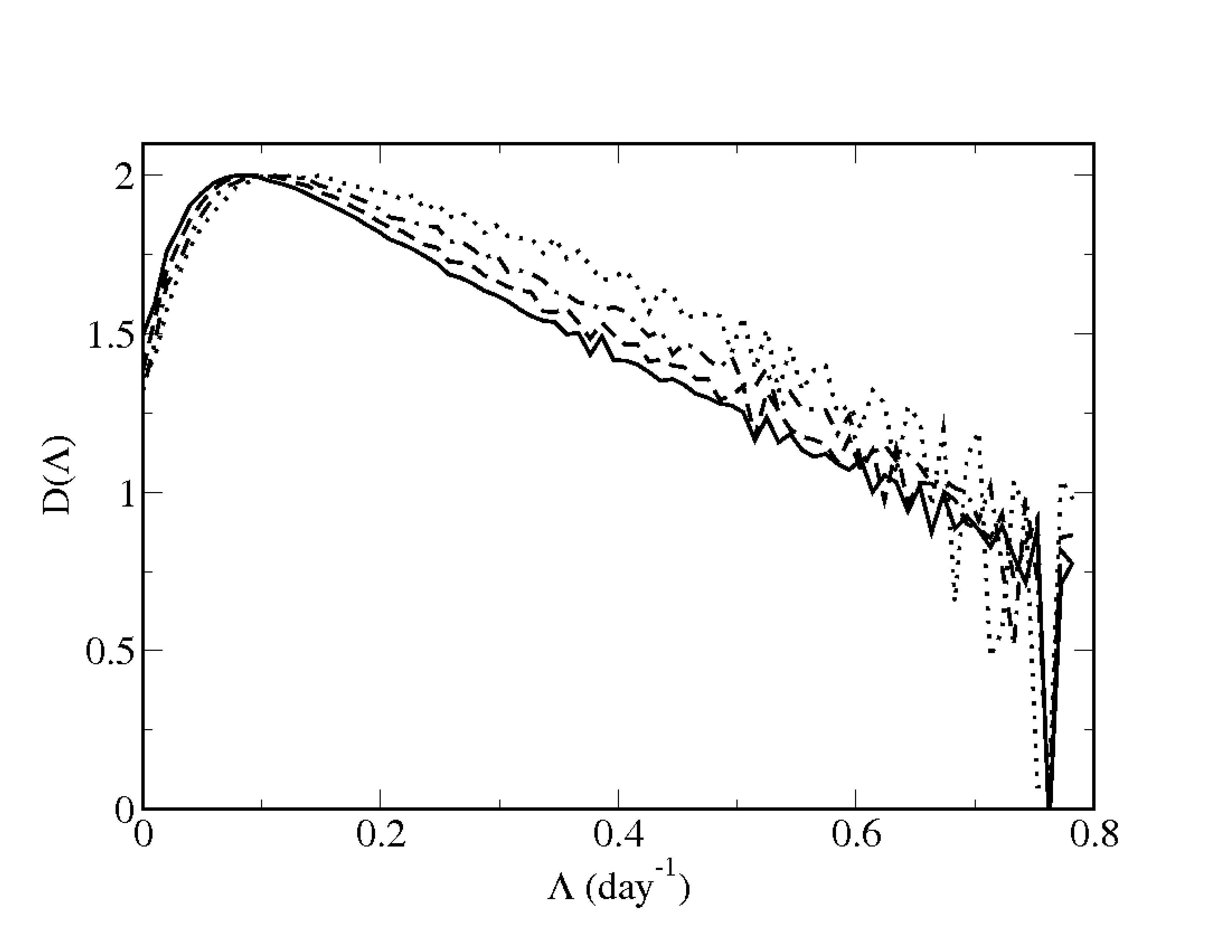}
\end{center}
\caption{{\bf Top:} Comparison of the probability density functions
$P(\delta_0, \Lambda)$ for the FSLEs at different resolutions. It
is obtained from the temporal average ($30$ snapshots) of
histograms. Dotted line is for $\delta_0=1/8 \degree$,
dashed-dotted line $\delta_0=1/16 \degree$, dashed $\delta_0=1/32
\degree$, and solid line for $\delta_0=1/64 \degree$.
{\bf Bottom:} $D(\Lambda)$ for different values of $\delta_0$. Dotted for
$\delta_0=1/8 \degree$, dashed-dotted line $\delta_0=1/16
\degree$, dashed $\delta_0=1/32 \degree$, and solid line line for
$\delta_0=1/64 \degree$.
 }
\label{fig:Dlambda}
\end{figure}

 In
Figure~\ref{fig:Dlambda}, top,  we show the histograms
(averaged over $30$ snapshots distributed among $15$ months)
normalized to have the same unitary are. One can see that when the resolution
gets finer the $P(\delta_0, \Lambda)$ narrows, and the peak height
increases, in agreement with Equation~(\ref{eq:FSLEpdf}). According to
the definition of Microcanonical Multifractals
\cite{Isern.2007,Turiel.2008}, the system will be multifractal if the
curves $D(\Lambda)$ estimated at different resolutions $\delta_0$
using Equation~(\ref{eq:Dlambda}) are all equal. This can be observed
in Figure~\ref{fig:Dlambda}, bottom. The collapse of the four curves
is not perfect due to the lost of translational invariance
produced by the small size of the domain -the
Balearic basin-  that we analyze. Thus, the interfaces of constant
$\Lambda$ values build an approximate
multifractal hierarchy and generalized scale invariance
is present in the FSLE field.

\section{Change of resolution of the velocity grid.}
\label{Ap:resolution}

In order to reduce the resolution of a given velocity grid, the easiest
way would be by subsampling the existing grid points or by
block-averaging the values of the velocity and assign the result to
the central grid point. However, in the case of a complex boundary such 
as the Mediterranean coast such a strategy is strongly inconvenient, as the
coarsening of the grid would imply to change land circulation
barriers (islands, straits). The disappearance of a land barrier or
the creation of a new one as a consequence of the coarsening would
imply a dramatic change in the value of FSLEs at all points affected
by the modified circulation; if the circulation patterns are rather
complex almost every point could be affected. We have thus preferred
to smooth the  velocity with a convolution kernel weighted with a
local normalization factor, and keeping the original resolution for
the data so land barriers are equally well described than in the original data.

We define the coarsening kernel of scale factor $s$, $\kappa_s$, as:

\begin{equation}
\kappa_s(x,y)\; =\; e^{-\frac{x^2+y^2}{2s^2}}
\end{equation}

\noindent
We disregard the introduction of a normalization factor at this point
since we will need to normalize locally later. The coarsened version of
the velocity vector would hence be given by the convolution of this
kernel with the velocity, denoted by $\kappa_s \star
\vec{v}$. A coarsening convolution kernel turns out to be convenient
with almost horizontal fields, as the derivatives commute with the
convolution operator so if $\nabla \vec{v}\approx  0$ hence $\kappa_s
\star \nabla \vec{v}\approx 0$. However, this coarsening scheme needs to
be improved. By convention, we take the velocity $\vec{v}$ as zero over
land points. For that reason, a simple convolution does not produce a
correct coarsened version of the velocity because points close to land would
experience a loss of energy.
The easiest way to correct this is to
normalize by the weight of the sea points. Let us first define the sea
mask $M(x,y)$ as $1$ over the sea and $0$ over the land. The normalization
weight is given by $\kappa_s \star M$. For points very far from the
land, this weight is just the normalization of $\kappa_s$. For points
surrounded by land points the weight takes the contributions from sea
points only. We thus define the coarsened version by a scale factor
$s$ of the velocity, $\vec{v}_s$, as:

\begin{equation}
\vec{v}_s \; =\; \frac{\kappa_s \star \vec{v}}{\kappa_s \star M}
\end{equation}

\begin{figure}[htb]
\begin{center}
\leavevmode
\includegraphics[width=14cm]{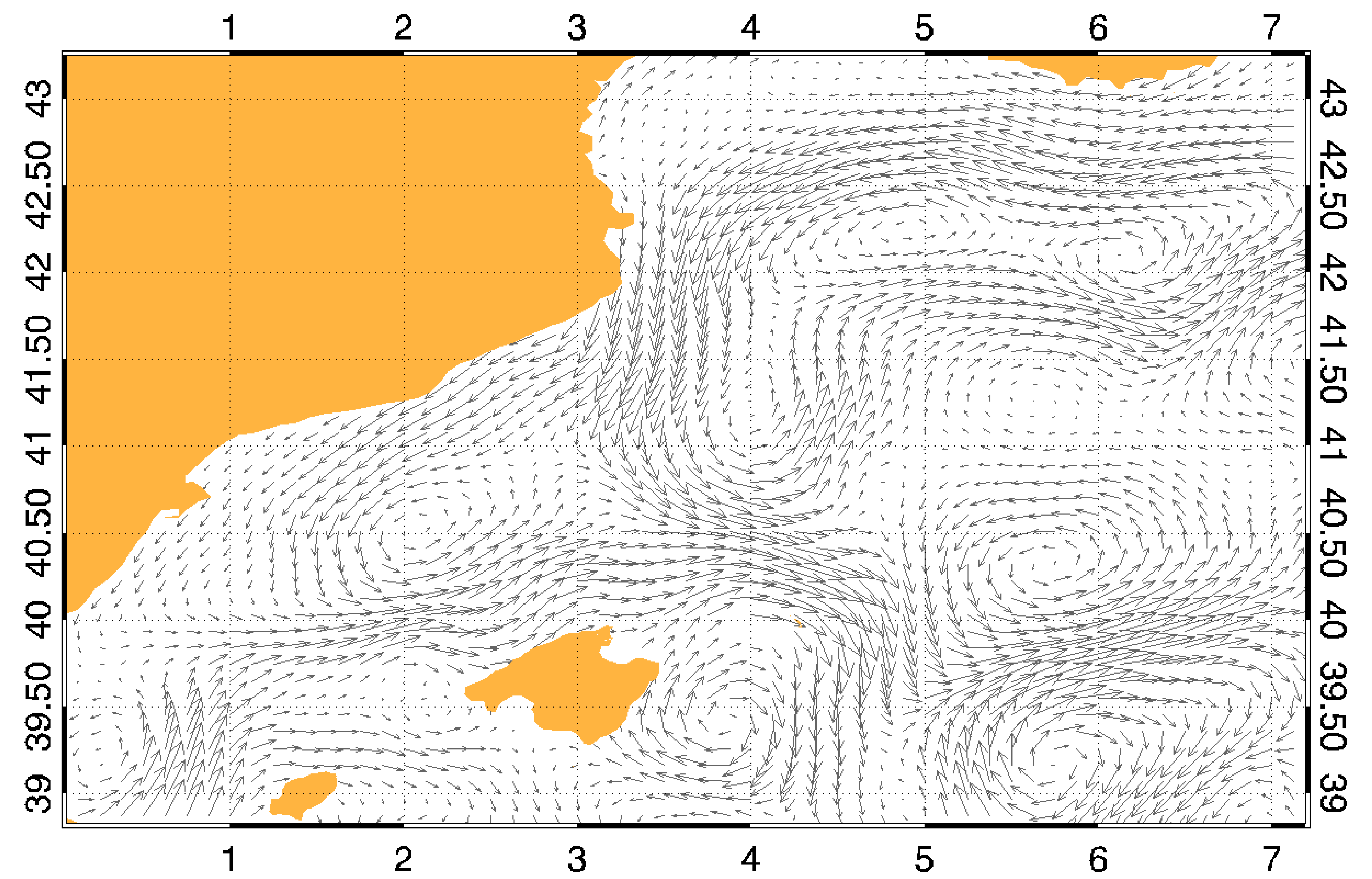}
\\
\leavevmode
\includegraphics[width=14cm]{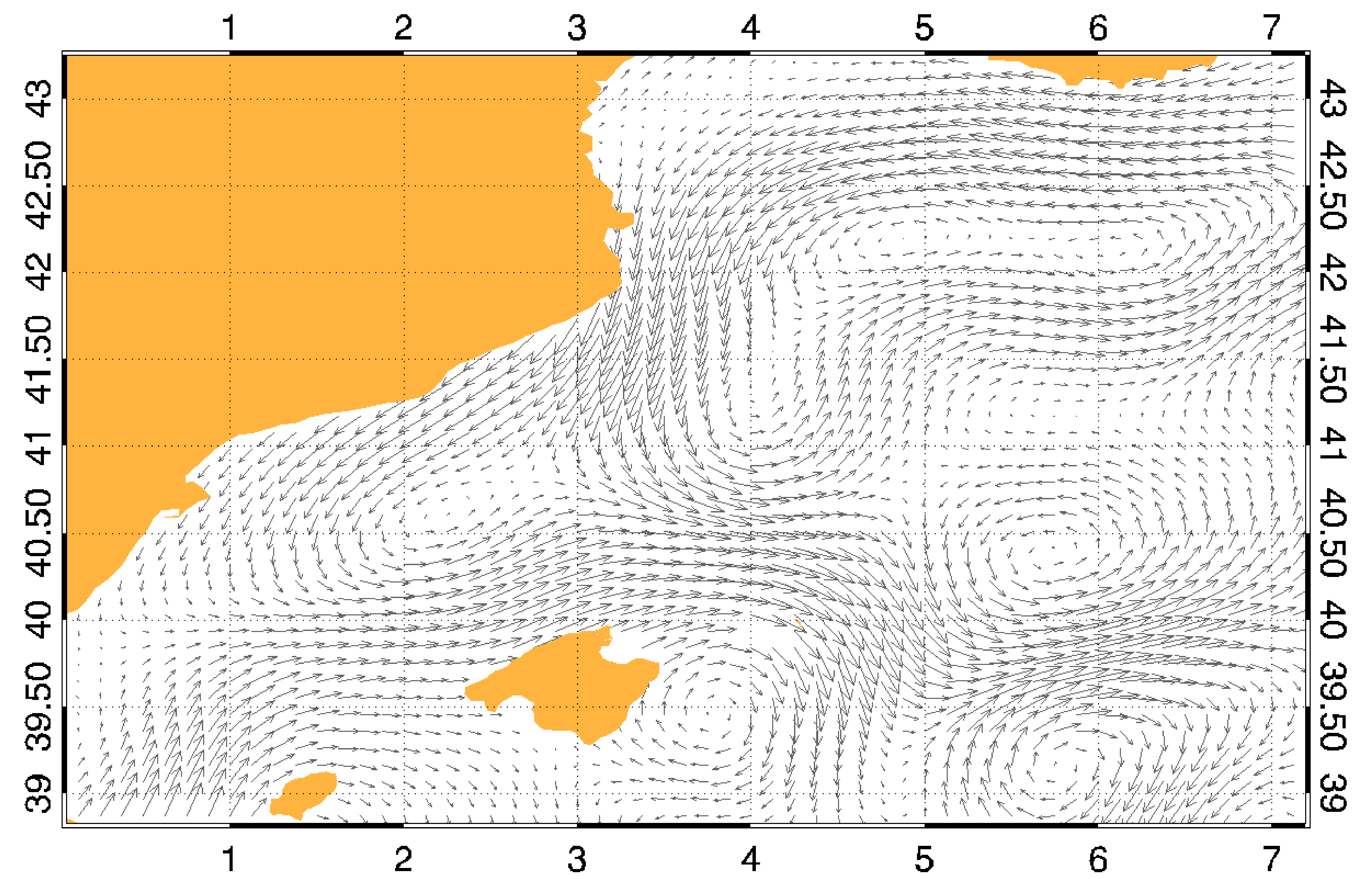}
\end{center}
\caption{{\bf Top:} Velocity field coarsened by a scale factor $s=2$,
  for a equivalent resolution $\delta_0=1/4 \degree$.
{\bf Bottom:} Velocity field coarsened by a scale factor $s=4$,
  for a equivalent resolution $\delta_0=1/2 \degree$.
 }
\label{fig:velocityres}
\end{figure}

In Figure~\ref{fig:velocityres} we show two examples of
coarsened velocities. We can see that typical circulation patterns are
coarsened as $s$ increases, while land obstacles are preserved. In
fact, if $\Delta_0$ is the velocity resolution scale, the effective
resolution scale of $\vec{v}_s$ is $s \Delta_0$ (the nominal
resolution, on the contrary, is the original one, $\Delta_0$).

\section{Introducing noise in the particle's trajectories.}
\label{Ap:noise}

In order to introduce noise in the particle's trajectories we resolve the following
system of stochastic equations:
\begin{eqnarray}
\frac{d\phi}{dt}=\frac{u(\phi,\lambda,t)}{R \cos(\lambda)}
+\frac{\sqrt{2D}}{R \cos(\lambda)}\xi_1(t),
\label{eq:noise1}
\\
\frac{d\lambda}{dt}=\frac{v(\phi,\lambda,t)}{R
}+\frac{\sqrt{2D}\xi_2(t)}{R }.
\label{eq:noise2}
\end{eqnarray}
 $\xi_i (t)\ \ i=1,2$ are the components of a two-dimensional Gaussian
white noise with zero mean and correlations $<\xi_i (t) \xi_j
(t')> =\delta_{ij}\delta(t-t')$.Eqs. (\ref{eq:noise1} and \ref{eq:noise2})
use a simple
white noise added to the trajectories. A more realistic
representation of small-scale Lagrangian dispersion in turbulent
fields requires using other kinds of correlated noises
\cite{Griffa.1996} but, as we are interested in examining
influences of the missing scales, it is convenient to use white
noise, since this would represent the extreme case of very
irregular trajectories which gives an upper bound to the effects of
more realistic smoother small scales. Thus the tests presented here
are similar to the ones considered before (perturbation of the velocity)
when adding uncorrelated perturbations to
the velocity, but here the perturbation acts at arbitrarily small
scales, as appropriate for a turbulent field, instead of being
smooth below a cutoff scale, as appropriate for modelling
observational errors.


\begin{thebibliography}{10}

\bibitem{Artale.1997}
V.~Artale, G.~Boffetta, A.~Celani, M.~Cencini, and A.~Vulpiani.
\newblock Dispersion of passive tracers in closed basins: Beyond the diffusion
  coefficient.
\newblock {\em Physics of Fluids}, 9:3162--3171, 1997.

\bibitem{Aurell.1997}
E.~Aurell, G.~Boffetta, A.~Crisanti, G.~Palading, and A.~Vulpiani.
\newblock Predictability in the large: an extension of the concept of
  {L}yapunov exponent.
\newblock {\em Journal of Physics A}, 30:1--26, 1997.

\bibitem{Beron-Vera.2008}
F.J. Beron-Vera, M.J. Olascoaga, and G.J. Goni.
\newblock Oceanic mesoscale eddies as revealed by {L}agrangian coherent
  structures.
\newblock {\em Geophysical Research Letters}, 35:L12603, 2008.

\bibitem{Boffetta.2001}
G.~Boffetta, G.~Lacorata., G.~Redaelli, and A.~Vulpiani.
\newblock Detecting barriers to transport: a review of different techniques.
\newblock {\em Physica D}, 159:58--70, 2001.

\bibitem{Buffoni.1997}
G.~Buffoni, P.~Falco, A.~Griffa, and E.~Zambianchi.
\newblock Dispersion processes and residence times in a semi-enclosed basin
  with recirculating gyres: {A}n application to the {T}yrrhenian sea.
\newblock {\em Journal of Geophysical Research - Oceans}, 102:18699--18713,
  1997.

\bibitem{dOvidio.2004}
F.~d'Ovidio, V.~Fern\'andez, E.~Hern\'andez-Garc\'ia, and C.~L\'opez.
\newblock Mixing structures in the mediterranean sea from finite-size lyapunov
  exponents.
\newblock {\em Geophysical Research Letters}, 31:L17203, 2004.

\bibitem{dOvidio.2009}
F.~d'Ovidio, J.~Isern-Fontanet, C.~L\'opez, E.~Hern\'andez-Garc\'{\i}a, and
  E.~Garc\'{\i}a-Ladona.
\newblock Comparison between {E}ulerian diagnostics and {F}inite-{S}ize
  {L}yapunov {E}xponents computed from altimetry in the {A}lgerian basin.
\newblock {\em Deep-Sea Research I}, 56:15--31, 2009.

\bibitem{Falconer.1990}
K.~Falconer.
\newblock {\em Fractal Geometry: Mathematical Foundations and Applications}.
\newblock John Wiley and sons, Chichester, 1990.

\bibitem{Fernandez.2005}
V.~Fern{\'a}ndez, D.E. Dietrich, R.L. Haney, and J.~Tintor{\'e}.
\newblock Mesoscale, seasonal and interannual variability in the
  {Mediterranean} sea using a numerical ocean model.
\newblock {\em Progress in Oceanography}, 66:321--340, 2005.

\bibitem{Frisch.1995}
U.~Frisch.
\newblock {\em Turbulence: The legacy of A.N. Kolmogorov}.
\newblock Cambridge Univ. Press, Cambridge MA, 1995.

\bibitem{Garcia-Olivares.2007}
A.~Garcia-Olivares, J.~Isern-Fontanet, and E.~Garcia-Ladona.
\newblock Dispersion of passive tracers and finite-scale {L}yapunov exponents
  in the {W}estern {M}editerranean {S}ea.
\newblock {\em Deep Sea Research I}, 54:253--268, 2007.

\bibitem{Griffa.1996}
A.~Griffa.
\newblock Applications of stochastic particle models to oceanographic problems.
\newblock In R.~Adler, P.~Muller, and B.~Rozovskii, editors, {\em Stochastic
  modelling in physical oceanography}. Birkhauser, 1996.

\bibitem{Haller.2000}
G.~Haller.
\newblock Finding finite-time invariant manifolds in two-dimensional velocity
  fields.
\newblock {\em Chaos}, 10(1):99--108, 2000.

\bibitem{Haller.1998}
G.~Haller and A.~Poje.
\newblock Finite time transport in aperiodic flows.
\newblock {\em Physica D}, 119:352--380, 1998.

\bibitem{Haller.2000b}
G.~Haller and G.~Yuan.
\newblock Lagrangian coherent structures and mixing in two-dimensional
  turbulence.
\newblock {\em Physica D}, 147:352--370, 2000.

\bibitem{Haza.2008}
A.~C. Haza, A.~C. Poje, T.~M. \"Ozg\"okmena, and P.~Martin.
\newblock Relative dispersion from a high-resolution coastal model of the
  {A}driatic {S}ea.
\newblock {\em Ocean Modelling}, 22:48--65, 2008.

\bibitem{Isern.2007}
J.~Isern-Fontanet, A.~Turiel, E.~Garcia-Ladona, and J.~Font.
\newblock Microcanonical multifractal formalism: application to the estimation
  of ocean surface velocities.
\newblock {\em Journal of Geophysical Research}, 112:C05{0}24, 2007.


\bibitem{Iudicone.2002}
D. Iudicone, G. Lacorata, V. Rupolo, R. Santoleri, and A. Vulpiani
\newblock Sensitivity of numerical tracer trajectories
to uncertainties in OGCM velocity fields.
\newblock {\em Ocean Modelling}, 4:313-325, 2002.


\bibitem{Joseph.2002}
B.~Joseph and B.~Legras.
\newblock Relation between kinematic boundaries, stirring, and barriers for the
  antartic polar vortex.
\newblock {\em Journal of Atmospheric Sciences}, 59:1198--1212, 2002.

\bibitem{Koh.2002}
T.-Y. Koh and B.~Legras.
\newblock Hyperbolic lines and the stratospheric polar vortex.
\newblock {\em Chaos}, 12(2):382--394, 2002.

\bibitem{Lapeyre.2002}
G.~Lapeyre.
\newblock Characterization of finite-time {L}yapunov exponents and vectors in
  two-dimensional turbulence.
\newblock {\em Chaos}, 12(2):688--698, 2002.


\bibitem{Mancho.2006}
A.M. Mancho, D.~Small, and S.~Wiggins.
\newblock A tutorial on dynamical systems concepts applied to lagrangian
  transport in oceanic flows defined as finite time data sets: theoretical and
  computational studies.
\newblock {\em Physics Reports}, 437:55--124, 2006.

\bibitem{Mancho.2008}
A.M. Mancho, E.~Hern\'andez-Garc\'{\i}a, D.~Small, S.~Wiggins, and
  V.~Fern\'andez.
\newblock Lagrangian transport through an ocean front in the {N}orth-{W}estern
  {M}editerranean {S}ea.
\newblock {\em Journal of Physical Oceanography}, 38:1222--1237, 2008.


\bibitem{Molcard.2006}
A. Molcard, A.C. Poje, and T. M. Ozgokmen.
\newblock
Directed drifter launch strategies for Lagrangian data assimilation using hyperbolic trajectories 
\newblock {\em Ocean Modelling}, 12: 268-289, 2006.


\bibitem{Okubo.1971}
A.~Okubo.
\newblock Ocean diffusion diagramas.
\newblock {\em Deep Sea Research}, 18:789--802, 1971.

\bibitem{Ott.1993}
E.~Ott.
\newblock {\em Chaos in Dynamical Systems}.
\newblock Cambridge Univ. Press, Cambridge (UK), 1993.

\bibitem{Chaos.2010}
T.~Peacock and J.~Dabiri.
\newblock Introduction to focus issue: Lagrangian coherent structures.
\newblock {\em Chaos}, 20:017501, 2010.

\bibitem{Poje.2010}
A.C. Poje, A.~C. Haza, T.~M. Ozgokme, M.~G. Magaldi, and Z.~D. Garraffo.
\newblock Resolution dependent relative dispersion statistics in a hierarchy of
  ocean models.
\newblock {\em Ocean Modelling}, 31:36--50, 2010.

\bibitem{Schneider.2005a}
J.~Schneider, V.~Fern\'andez, and E.~Hern\'andez-Garc\'ia.
\newblock Leaking method approach to surface transport in the {Mediterranean
  Sea} from a numerical ocean model.
\newblock {\em Journal of Marine Systems}, 57:111--126, 2005.



\bibitem{TewKai.2009}
E.~Tew Kai, V.~Rossi, J.~Sudre, H~Weimerskirch, C.~L\'opez,
  E.~Hern\'andez-Garc\'\i a, F.~Marsac, and V.~Garcon.
\newblock Top marine predators track lagrangian coherent structures.
\newblock {\em Proceedings of the National Academy of Sciencies of the USA},
  106:8245--8250, 2009.





\bibitem{Turiel.2002}
A.~Turiel and A.~del Pozo.
\newblock Reconstructing images from their most singular fractal manifold.
\newblock {\em IEEE Trans. Im. Proc.}, 11:345--350, 2002.

\bibitem{Turiel.2006}
A.~Turiel, C.~P\'erez-Vicente, and J.~Grazzini.
\newblock Numerical methods for the estimation of multifractal singularity
  spectra on sampled data: a comparative study.
\newblock {\em Journal of Computational Physics}, 216(1):362--390, July 2006.

\bibitem{Turiel.2008}
A.~Turiel, H.~Yahia, and C.~P\'erez-Vicente.
\newblock Microcanonical multifractal formalism: a geometrical approach to
  multifractal systems. {Part I}: Singularity analysis.
\newblock {\em Journal of Physics A}, 41:015501, 2008.

\end{thebibliography}

\end{document}